# Design of high-temperature f-block molecular nanomagnets through the control of vibration-induced spin relaxation

*L. Escalera-Moreno,[a] Jose J. Baldoví,*,[b] A. Gaita-Ariño,[a] E. Coronado[a]*

[a]Instituto de Ciencia Molecular (ICMol), Universidad de Valencia, c/ Catedrático José Beltrán 2, 46980, Paterna, Spain

[b]Max Planck Institute for the Structure and Dynamics of Matter, Luruper Chaussee 149, 22761, Hamburg, Germany

**Corresponding Author**

*E-mail: jose.baldovi@mpsd.mpg.de

ABSTRACT. One of the main roadblocks that still hampers the practical use of molecular nanomagnets is their cryogenic working temperature. In the pursuit of rational strategies to design new molecular nanomagnets with increasing blocking temperature, *ab initio* methodologies play an important role by guiding synthetic efforts at the lab stage. Nevertheless, when evaluating vibration-induced spin relaxation, these methodologies are still far from being computationally fast enough to provide a useful predictive framework. Herein, we present an inexpensive first-principles method devoted to evaluating vibration-induced spin relaxation in molecular f-block single-ion magnets, with the important advantage of requiring only one CASSCF calculation. We use a case study to illustrate the method, and propose chemical modifications in the ligand environment with the aim of suppressing spin relaxation.

**TOC GRAPHICS**

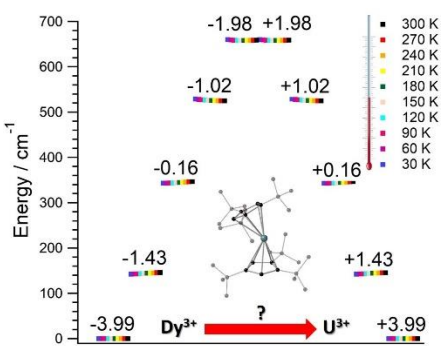

**KEYWORDS** Coordination Chemistry, Molecular Nanomagnet, Single-Ion Magnet, Molecular Magnetic Relaxation, Molecular Vibration, Computational Quantum Chemistry.

Molecular nanomagnets are magnetic molecules where a bistable ground state characterized by a hysteresis loop results in a memory effect, which is harnessed for classical information storage. These nanomagnets can contain either a single magnetic metal ion, known as Single-Ion Magnets (SIMs);[1] or more than one, called Single-Molecule Magnets (SMMs).[2] Two key parameters characterize the performance of a molecular nanomagnet, namely, magnetic relaxation time and blocking temperature. The former is the timescale where molecular nanomagnets preserve classical information at a given temperature, and the latter gives the maximum temperature that allows observing magnetic hysteresis.

The field of molecular nanomagnets is at a crucial point. Over the last few years, we have witnessed the discovery of new molecular nanomagnets that have allowed an outstanding increase in the blocking temperature, first from 30 K to 60 K (2017),[3,4] and then up to 80 K (2018).[5] This trend is drawing an unprecedented interest by many researchers around the world, and demands urgent attention from the theoretical side. Indeed, the large rise of blocking temperatures opens the possibility to incorporate molecular nanomagnets in devices operating at higher temperatures. But first, we need to deepen the understanding of those magnet-behavior destroying processes, known as spin relaxation.

The target features that have commonly been addressed to block spin relaxation are (i) the ground electron spin quantum number $J$, and (ii) the barrier height that separates the two components of the bistable ground state.[6] Indeed, the search for new nanomagnets by increasing (i) and (ii) is consistent with an Orbach-like relaxation mechanism, which drives the spin population across the barrier. While this strategy has worked for so long,[7,8] the recent interest in molecular nanomagnets operating at high temperatures, where spin-vibration coupling dominates relaxation, makes this picture be insufficient.[6] Hence, to gain control on relaxation at increasing temperatures, spin-vibration coupling must be incorporated in the theoretical methods.

The current pursuit of predictive power is encouraging the development of fully *ab initio* methodologies.[9,10] Nevertheless, first-principles evaluations of spin-vibration coupling are known to be computationally demanding,[3,9,10] This fact constitutes a crucial limitation that makes state-of-the-art *ab initio* methods be impractical when guiding efforts at the lab stage. Thus, searching for new methodologies able to circumvent this computational bottleneck is of paramount urgency. In the case of lanthanide-based SIMs, there already exist affordable semi-empirical approximations devoted to determining the electronic structure,[11,12] whose accuracy can become comparable to that of *ab initio* calculations.[13,14]

Herein, we present an inexpensive first-principles method devoted to lanthanide and uranium SIMs, with the aim of evaluating vibration-induced spin relaxation. It allows estimating relaxation effective barriers $U_{eff}$, relaxation pathways and relaxation times $\tau$ at any temperature. Crystal field parameters (CFPs) are determined by millisecond calculations, and only one CASSCF evaluation is required. The method identifies those vibrations promoting relaxation in order to re-design the given molecule, and incorporates for the first time a temperature dependence in the spin-vibration coupling matrix elements. Contributions from spin-spin dipole coupling to $U_{eff}$ and $\tau$ can be incorporated by resorting to recent first-principles models.[15] Since the barrier height may increase from lanthanides to actinides due to a stronger ligand field, and given the challenging computational nature of the $U^{3+}$ ion,[16,17] we propose to evaluate the effectiveness of bis-metallocenium ligands on actinides and test the efficiency of our method on the hypothetical analog $[U(Cp^{ttt})_2]^+$ of $[Dy(Cp^{ttt})_2]^+$, which holds one of the latest records in the blocking temperature.[3] The method consists of the following steps:

**Step 1**. The relevant atom set is relaxed until reaching a minimum in its potential energy surface.[6] This set may be the SIM itself,[9] or the unit cell of a crystal containing the SIM.[10] After calculating the vibrational spectrum, we save harmonic frequencies $\{v_j\}_j$, reduced masses $\{m_j\}_j$, and displacement vectors $\{\vec{w_j}\}_j$. These determine the 3D-direction in which each atom vibrates around its equilibrium position.

**Step 2**. One performs a CASSCF calculation on the SIM experimental structure to extract the lowest $2J+1$ energies, where $J$ is the metal ion ground electron spin quantum number. Then, once the coordinate origin is placed at the metal experimental position, the experimental positions of only the metal-coordinating atoms are introduced in the code SIMPRE, see SI.[11,12] This code first calculates the CFPs by considering each coordinating atom as an effective point charge, and then performs a millisecond diagonalization of the ground $J$ Crystal Field Hamiltonian. The charge magnitudes and the metal-charge radial distances are varied to fit the CASSCF energies,[18,19] see SI. Thus, we project the CASSCF information onto the first coordination sphere via effective parameters. Note that the contribution of the coordinating atoms to the ligand field almost recover the whole effect of magnetic anisotropy. Nevertheless, one can include non-coordinating ligand atoms if a significant contribution is expected. To reduce computational cost, quite often it will be enough to keep the same charge magnitude and radial distance variation in each coordinating atom.

This procedure is fully *ab initio*, but one can avoid the CASSCF evaluation and use the experimental energies if they are available. In this case, the experimental structure used in SIMPRE should be determined at the same temperature as that of the experimental energies. Now, the coordinating atom positions of the relaxed geometry are radially varied with the same fitting distance variations determined in SIMPRE. By using the same found charge values, SIMPRE calculates the equilibrium CFPs $\left\{\left(A_k^q \langle r^k \rangle\right)_{eq}\right\}_{k,q}$ in Stevens notation. Unlike lanthanides, excited states beyond the ground $J$ multiplet may also influence the low-lying electronic structure of actinide SIMs. Thus, in case of $U^{3+}$, to determine the charge magnitude and the radial distance variation, the energy fitting must be replaced by a fitting of the SIMPRE CFPs to the CFPs either CASSCF or experimental.

The diagonalization in SIMPRE of the equilibrium Crystal Field Hamiltonian $\hat{H}_{eq} = \sum_{k=2,4,6}\sum_{q=-k}^{k} \left(A_k^q \langle r^k \rangle\right)_{eq} \eta_k \hat{O}_k^q$, where $\eta_k$ are the Stevens coefficients and $\hat{O}_k^q$ are the Stevens equivalent operators,[11,12] provides the lowest $2J+1$ equilibrium eigenstates and energies, see SI. For $U^{3+}$, the diagonalization is performed in the code CONDON,[20] which contains the excited states beyond the ground $J$ multiplet, and the CFPs must be introduced in Wybourne notation. These eigenstates are truncated to ordered basis set $\{|-J\rangle,...,|+J\rangle\}$ of the ground $J$ multiplet and then renormalized.

The perturbing Hamiltonians $\left\{\hat{H}_j = \sum_{k=2,4,6}\sum_{q=-k}^{k} \Delta\left(A_k^q \langle r^k \rangle\right)_j (T) \eta_k \hat{O}_k^q \right\}_j$, which are also built in the above ordered basis set, account for the perturbation to the equilibrium electronic structure from each vibrational mode $j$, see SI. Their determination requires to estimate the temperature-dependent change $\Delta\left(A_k^q \langle r^k \rangle\right)_j (T)$ produced in $\left(A_k^q \langle r^k \rangle\right)_{eq}$ after activating each mode $j$. We use a model derived by us elsewhere,[9] which provides the following perturbative expression up to second-order in mode coordinate $Q_j$:

$$\Delta\left(A_k^q \langle r^k \rangle\right)_j (T) = \frac{\hbar}{4\pi} \left(\frac{\partial^2 A_k^q \langle r^k \rangle}{\partial Q_j^2}\right)_{eq} \frac{1}{m_j v_j} \left(\langle n_j \rangle + \frac{1}{2}\right) \quad \text{Eq. 1}$$

Thus, each $\hat{H}_j$ allows determining the spin-vibration coupling matrix element $\langle i|\hat{H}_j|f\rangle$, see SI. The temperature dependence is introduced for the first time in these elements through each boson number $\langle n_j \rangle = 1/\left(e^{h\nu_j/k_B T} - 1\right)$. The procedure to calculate the derivatives $\left(\partial^2 A_k^q \langle r^k \rangle / \partial Q_j^2\right)_{eq}$ is found in SI.

**Step 3**. This step is undertaken by solving the master equation, Eq. 2,[3,21,22] which describes the time evolution of the spin population across the lowest $2J+1$ equilibrium eigenstates. The energy that induces the spin to relax comes from the coupling with surrounding vibrations. Intuitively, at each time $t$ there is a probability $p_i(t)$ of being in an eigenstate $|i\rangle$. At a time $t+dt$ the system may make a transition to a different eigenstate $|f\rangle$ with a probability $\gamma_{if} dt$, either by absorbing or by emitting a vibration quantum. The net difference between the incoming $\gamma_{fi} p_i$ and outcoming $\gamma_{if} p_f$ spin populations equals the change in time of $p_i$. Thus, once the transitions are assumed to be independent, these probabilities evolve as:

$$\frac{dp_i(t)}{dt} = \sum_{f=1, f \neq i}^{2J+1} \left[\gamma_{if} p_f(t) - \gamma_{fi} p_i(t)\right], \quad i=1,...,2J+1 \quad \text{Eq. 2}$$

The transition rates $\gamma_{if}$ and $\gamma_{fi}$ account for the spin population flow between $|i\rangle$ and $|f\rangle$, and their expressions depend on the relaxation process to model. We include two important processes: (i) Orbach and (ii) second-order Raman, see SI. The determination of the most likely relaxation pathway provides further insight on relaxation. This allows identifying the vibrations that promote each relaxation step, and modifications on the molecular structure can then be proposed to suppress relaxation. All details are found in SI.[3,21-23] The same procedure applies if the Crystal Field Hamiltonians are expanded to include a static magnetic field via a Zeeman term.

**A case study: [U(Cp$^{ttt}$)$_2$]$^+$.** Because of the larger size of U$^{3+}$ compared to Dy$^{3+}$, we propose to use a counter-ion different from [B(C$_6$F$_5$)$_4$]$^-$ with bulkier substituents in the synthetic process, as F$^-$ could coordinate the U$^{3+}$ ion. Recently, it has been reported a uranium-based bis-metallocenium SIM,[24] which proves the possibility of synthesizing these uranium derivatives.

Since we lack a [U(Cp$^{ttt}$)$_2$]$^+$ structure, we use the experimental geometry of [Dy(Cp$^{ttt}$)$_2$]$^+$.[3] By replacing Dy$^{3+}$ by U$^{3+}$, we carry out the geometry relaxation and the vibrational spectrum calculation, Fig.1 and SI. First, we should perform a CASSCF evaluation on a real structure of [U(Cp$^{ttt}$)$_2$]$^+$ to obtain $\left\{ \left( A_k^q \left\langle r^k \right\rangle \right)_{eq} \right\}_{k,q}$. Since this is not possible, we proceed as explained in SI to determine the equilibrium electronic structure, Fig. 1.

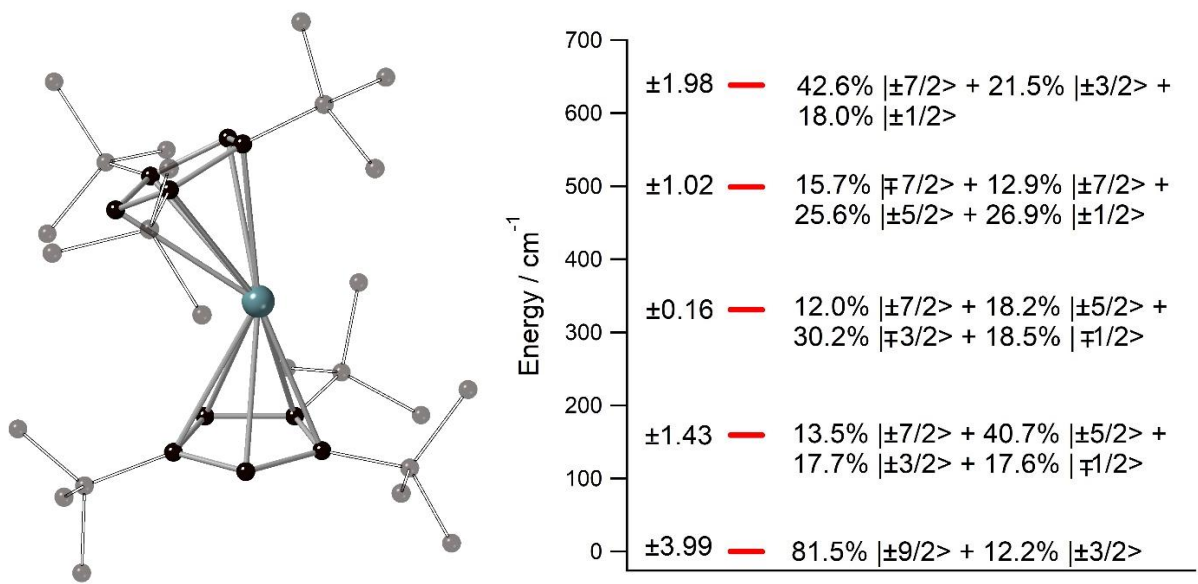

**Figure 1**. Left: [U(Cp$^{ttt}$)$_2$]$^+$ relaxed geometry. Hydrogen atoms omitted for clarity. Right: [U(Cp$^{ttt}$)$_2$]$^+$ lowest $2J+1=10$ equilibrium energies with wave-functions on the right (amplitudes < 10% not shown) and their $J_z$ operator expectation values on the left.

Once $\left(A_k^q \langle r^k \rangle\right)_{eq}$ and $\left(\partial^2 A_k^q \langle r^k \rangle / \partial Q_j^2\right)_{eq}$ are determined, we calculate the CFPs thermal evolution.[9] From Fig. 2, important contributions from off-diagonal CFPs are clearly observed, which leads to a sizeable $|m_J\rangle$ mixing in the equilibrium eigenstates, see SI. This fact opposes a good SIM behavior, where the diagonal CFPs should largely dominate over the off-diagonal ones. Another feature is the small yet significant relative thermal change in some CFPs. A proper tuning in the chemical structure aimed to reduce the variations $\Delta\left(A_k^q \langle r^k \rangle\right)_j$ would improve the molecular nanomagnet performance.[9] Indeed, this reduction would make the matrix elements $\langle i|\hat{H}_j|f\rangle$ and the transition rates $\gamma$ be smaller, since the perturbing Hamiltonians $\hat{H}_j$ are proportional to $\Delta\left(A_k^q \langle r^k \rangle\right)_j$.

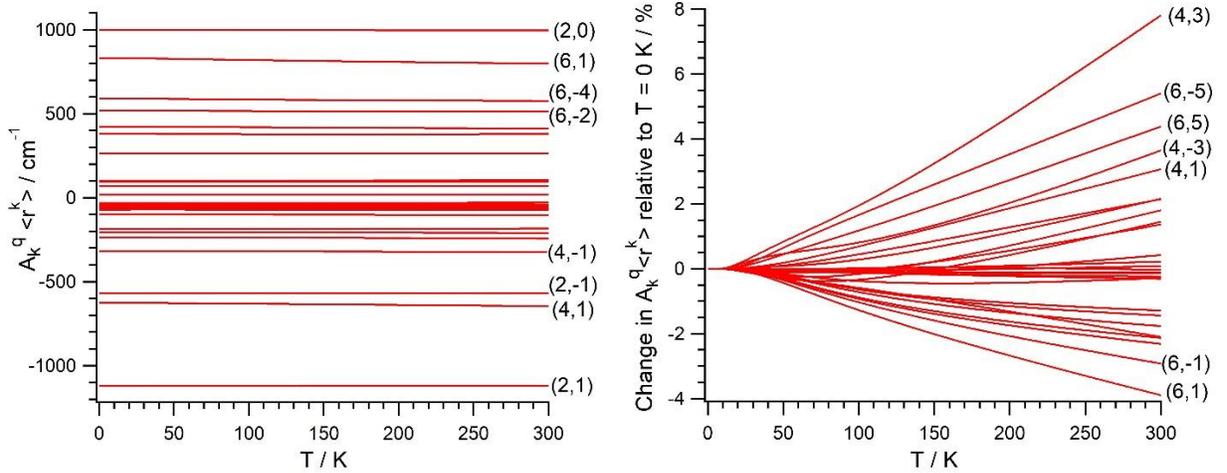

**Figure 2**. Absolute (left) and relative to T = 0 K (right) thermal evolution of the CFPs. Some parameters are identified as (k,q), where k and q are the scripts k = 2, 4, 6 and q = -k,…,+k.

Fig. 3 shows the thermal dependence of relaxation time $\tau$ when Orbach transitions rates are used in Eq. 2. Above 30 K, where there exists a high enough number of available phonons, the thermally-activated regime is at play and the spin population crosses the barrier through excited doublets, Fig. 4. Below 30 K, little or rather negligible spin population is promoted to the second excited doublet, which mostly tunnels the barrier through the first excited doublet. Nevertheless, the calculated Orbach-based relaxation times are now much larger than those found in most of SIMs. Thus, a different mechanism could be dominating spin relaxation such as quantum tunneling between the ground doublet components, which is not recovered by our approach but commonly observed at low temperatures.[4,5]

In this thermally-activated regime, the estimated effective barrier $U_{eff}$ = 292 K is in the hundreds of kelvin, something usual in a large set of molecular nanomagnets,[25] and is found around 40 cm$^{-1}$ above the first excited doublet in Fig. 1. The Orbach prefactor, $\tau_0 = e^{-17.43} = 2.7 \cdot 10^{-8}$ s, is within the usual range ($10^{-6} - 10^{-10}$ s) for SIMs with a barrier of comparable height. We also evaluated Eq. 2 with the second-order Raman transition rates. The Raman-based $\tau$ values, see Table S1, are much larger than the ones in Fig. 3. Thus, the second-order Raman process should be discarded as a competitive mechanism in an experiment, as found in $[Dy(Cp^{ttt})_2]^+$.[3]

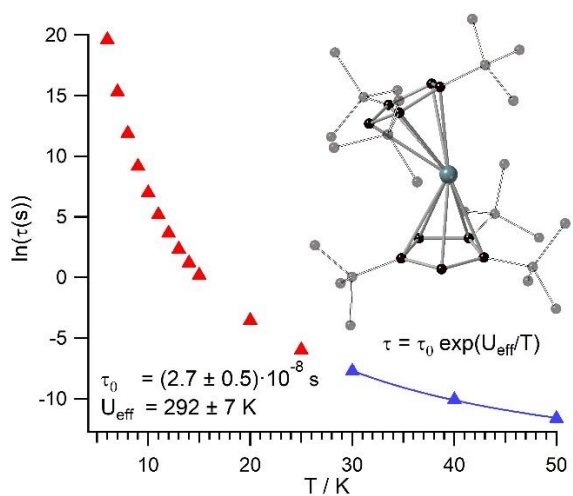

**Figure 3**. Thermal evolution of the Orbach-based relaxation time $\tau$, along with fit to determine both the Orbach prefactor $\tau_0$ and the effective barrier $U_{eff}$ in the thermally-activated regime ($T \geq 30K$).

Even employing bis-metallocenium ligands, known to offer a strong axial crystal field in Dy-based SIMS,[3-5] our calculated $[U(Cp^{ttt})_2]^+$ effective barrier ($\sim 292K$) is one order of magnitude below those reported for $[Dy(Cp^{ttt})_2]^+$ ($\sim 1760K$),[3] and Dy-5* ($\sim 2217K$).[5] Besides, the maximum temperature in $[U(Cp^{ttt})_2]^+$ ($\sim 50K$) at which the experimental relaxation time is still above the standard experimental detection limit is also clearly smaller compared to $[Dy(Cp^{ttt})_2]^+$ ($\sim 112K$),[3] and Dy-5* ($\sim 138K$).[5] As could be expected for a decrease of about one order of magnitude in the barrier height, the calculated $[U(Cp^{ttt})_2]^+$ Orbach prefactor $\tau_0 \sim 2.7 \cdot 10^{-8} s$ is three to four orders of magnitude above the ones corresponding to these two Dy-based SIMs ($\tau_0 \sim 2.0 \cdot 10^{-11} s$ and $\tau_0 \sim 4.2 \cdot 10^{-12} s$, resp.).[3,5]

Our $[U(Cp^{ttt})_2]^+$ Orbach prefactor is among the smallest ones that have been experimentally determined in uranium SIMs.[16,17] On the other hand, there do exist two significant advances respect to previous uranium SIMs: (i) the standard effective barrier is in the range of dozens of kelvin,[16,17] while $[U(Cp^{ttt})_2]^+$ reaches several hundreds of kelvin ($\sim 292K$), (ii) by assuming that the thermally-activated regime dominates between 30 K – 50 K in $[U(Cp^{ttt})_2]^+$, while it was not possible to determine relaxation times beyond ~5 K,[16,17] the experimental detection limit $\tau \sim 10^{-6} s$ in case of $[U(Cp^{ttt})_2]^+$ would be found at 50 K. Thus, it seems that $[U(Cp^{ttt})_2]^+$ is not expected to display hysteresis temperatures that compete with the ones of $[Dy(Cp^{ttt})_2]^+$ and Dy-5* SIMs. This is unsurprising since, after all, the $Cp^{ttt}$ rings were optimized for dysprosium, which may present somewhat different requirements in uranium. However, our methodology is efficient enough to offer a path forward in the rational design of ligands that result in uranium SIMs with optimized performance.

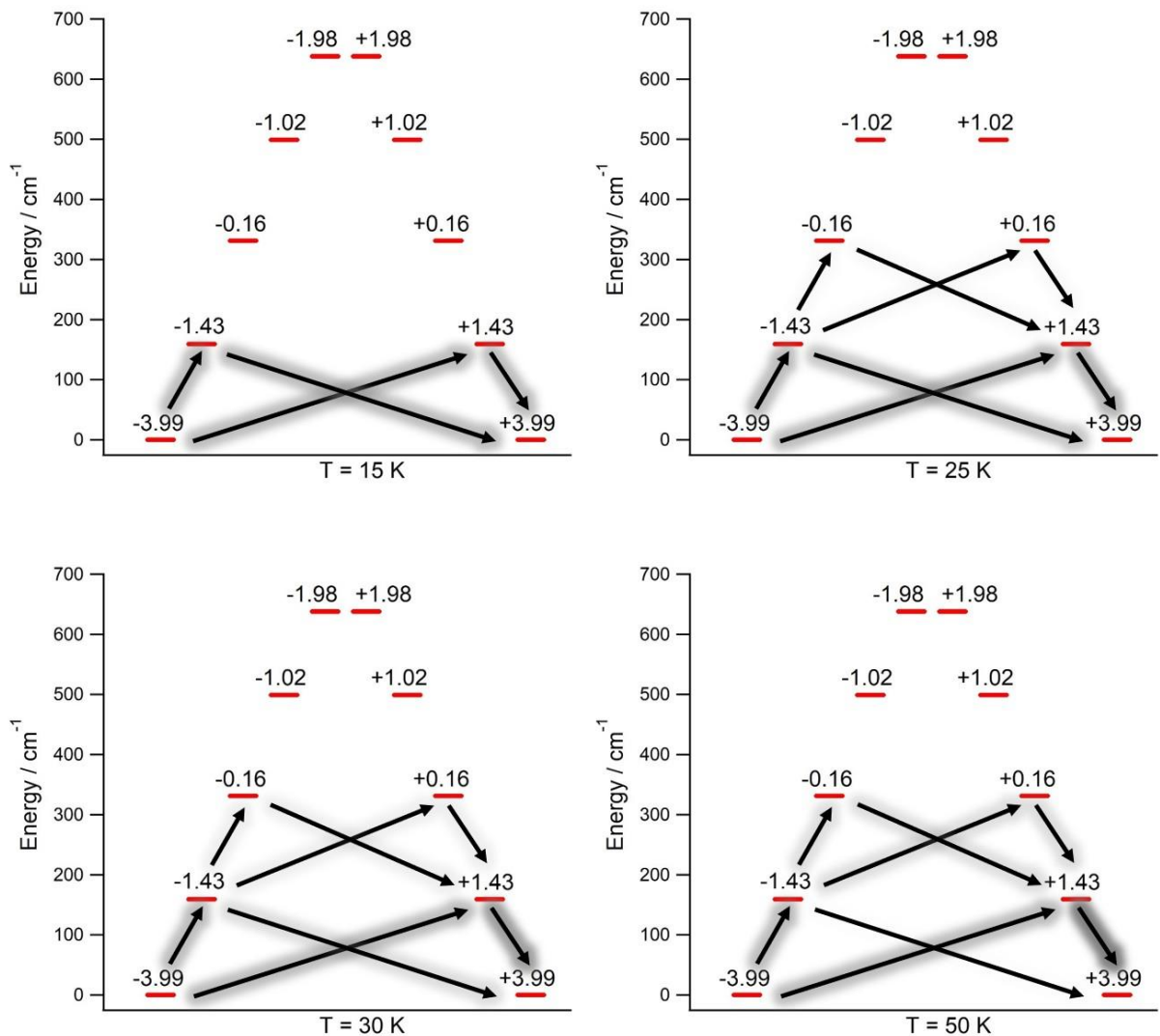

**Figure 4**. Orbach-driven relaxation pathways starting at the $\langle J_z \rangle = -3.99$ equilibrium eigenstate with unity population. Arrow shades are proportional to the percentage of the transient spin population, see SI. The outcome population sum from a given eigenstate equals the income population sum to the same eigenstate. Transient populations lesser than 1.0% not shown.

Let us now analyze the $[U(Cp^{ttt})_2]^+$ vibrations determining the transition rates that drive the relaxation pathways in Fig. 4, see animations and SI, and offer some strategies to reduce their detrimental effects. Two of them involve methyl rotations in the $Cp^{ttt}$ ring substituents. These rotations could be partially suppressed by replacing the methyl groups –$CH_3$ by the heavier fluorinated analogs –$CF_3$. A quick inspection shows that there exist similar vibrations promoting relaxation in $[Dy(Cp^{ttt})_2]^+$, Dy-5* and $[U(Cp^{ttt})_2]^+$. On one hand, rocking-like deformations of the $Cp^{ttt}$ rings where directly bounded hydrogen atoms move towards and away from the metal ion are present in $[Dy(Cp^{ttt})_2]^+$ and $[U(Cp^{ttt})_2]^+$. On the other hand, breathing vibrations where the two coordinating rings move towards and away from the metal ion are found in $[U(Cp^{ttt})_2]^+$ and Dy-5*. The rocking-like deformations in $[Dy(Cp^{ttt})_2]^+$, also found in $[U(Cp^{ttt})_2]^+$, were already blocked in ref. 5 by placing bulkier substituents in the coordinating rings. It worked as expected since both the effective barrier and the blocking temperature were increased respect to $[Dy(Cp^{ttt})_2]^+$. A possible strategy not yet explored to remove the breathing vibrations could be bounding the two coordinating rings, such as in stapled bis-phthalocyanines. Moreover, the frequencies of the $[U(Cp^{ttt})_2]^+$ detrimental vibrations, see SI, closely match the gaps between the equilibrium ground and first excited doublets (159.3 cm$^{-1}$), and first and second excited doublets (171.7 cm$^{-1}$), Fig. 1. Thus, the $[U(Cp^{ttt})_2]^+$ performance would also benefit from any structural modification that takes these frequencies out of resonance respect to these gaps.

All in all, the most important conclusions of this work are the following. We have proposed a novel first-principles methodology aimed to simulate the vibration-induced spin relaxation in f-block SIMs. The method offers the important advantage of drastically reducing the computation time while keeping the calculation accuracy inside an acceptable range. Indeed, all but one of the expensive CASSCF calculations required in previous methods are replaced by millisecond calculations. Besides, our approach introduces for the first time a temperature dependence in the spin-vibration coupling matrix elements. To demonstrate this methodology, we consider the highly-performing SIM [Dy(Cp$^{ttt}$)$_2$]$^+$ and find that the replacement of Dy$^{3+}$ by U$^{3+}$ does not result in an enhanced performance. Yet, [U(Cp$^{ttt}$)$_2$]$^+$ does seem to outperform all previously reported uranium SIMs. One of the critical factors that promote spin relaxation in [U(Cp$^{ttt}$)$_2$]$^+$ is the noticeable mixing among the $|m_J\rangle$ components in the equilibrium eigenstates. Importantly, this mixing is also found in previously reported uranium SIMs,[16,17,26] but not in the cutting-edge Dy-based SIMs [Dy(Cp$^{ttt}$)$_2$]$^+$ and Dy-5* even though the ligand coordination is not strictly axial.[3,5] Among those vibrations that promote spin relaxation, there are still atomic movements left to block. These involve methyl rotations and breathing deformations, which could be removed by fluorination and stapling the coordinating rings each other, resp. Hence, we conclude that there may be still room for further improvement in these bis-metallocenium-based uranium SIMs.

ASSOCIATED CONTENT

**Supporting Information**. The following files are available free of charge.

Supplementary information (file escalera-baldovi-coronado-SI). Input and outputs of the geometry relaxation and the vibrational spectrum calculation (subfolder Relax-Geom-Vib-Spec-Calc). CFPs evolution against distortion coordinates (subfolder CFPs-vs-Dist-Coord). Fortran codes, inputs and outputs to evaluate spin relaxation and to calculate CFPs thermal evolutions (Home-Made-Codes). Orbach-driven magnetization decays with time (subfolder Orbach-mag). Master matrices and diagonalization, relaxation pathways and vibrational mode contributions (subfolder Orbach-rp-Raman). Animations of the [U(Cp$^{ttt}$)$_2$]$^+$ calculated vibrational modes (subfolder Animations).

AUTHOR INFORMATION

**Notes**

The authors declare no competing financial interests.

ACKNOWLEDGMENT

The present work has been funded by the EU (ERC-2014-CoG-647301 DECRESIM, ERC-2018-AdG-788222 MOL-2D, COST Action CA15128 MOLSPIN, and the European Project SUMO of the call QuantERA), by the Spanish MINECO (grant CTQ2015-66223-C2-2-P, grant CTQ2017-89993 cofinanced by FEDER, grant MAT2017-89528, and Unit of excellence "María de Maeztu" MDM-2015-0538), and by the Generalitat Valenciana (Prometeo Program of Excellence). JJB acknowledges the EU for a Marie Curie Fellowship (H2020-MSCA-IF-2016-751047). AGA acknowledges funding by the MINECO (Ramón y Cajal Program).

# Supplementary Information

# Design of high-temperature f-block molecular nanomagnets through the control of vibration-induced spin relaxation

L. Escalera-Moreno,[a] Jose J. Baldoví,[*,b] A. Gaita-Ariño, E. Coronado[a]

a. Instituto de Ciencia Molecular (ICMol), Universidad de Valencia, c/ Catedrático José Beltrán 2, 46980, Paterna, Spain

b. Max Planck Institute for the Structure and Dynamics of Matter, Luruper Chaussee 149, 22761, Hamburg, Germany

*. Corresponding author

## Geometry relaxation and vibrational spectrum calculation

This step is commonly accomplished by methods based on Density Functional Theory, as they provide an acceptable accuracy-cost ratio. We use the package Gaussian09 to perform the geometry relaxation and the vibrational spectrum calculation.[1] In the folder Relax-Geom-Vib-Spec-Calc one finds the input uranocenium.inp and the output uranocenium.log which crashed and was subsequently restarted in uranocenium-r.log. In the output uranocenium-r.log one finds the relaxed geometry, the harmonic frequencies, the reduced masses, the force constants and the displacement vectors of each vibrational mode.

In the input uranocenium.inp one finds the experimental crystallographic geometry of $[Dy(Cp^{ttt})_2]^+$ (ref. 3 in main text), where we have replaced $Dy^{3+}$ by $U^{3+}$. The resultant structure has a charge of +1 and a ground spin multiplicity (2S+1) of 4. This structure is relaxed in vacuum. Of course, a more realistic relaxation often requires to include the nearest environment of the molecule, although it may be kept frozen in order not to increase the computational cost (ref. 9 in main text). Since our main goal in this manuscript is to focus on the novel methodology that we are proposing, we decided to relax this geometry in vacuum in order not to unnecessarily increase the computational cost. Thus, we apply our method to $[U(Cp^{ttt})_2]^+$ and study its magnetic relaxation mediated only by molecular vibrations in vacuum. As a matter of fact, the experimental geometry of $[Dy(Cp^{ttt})_2]^+$ was also relaxed in vacuum elsewhere by independent authors, and reasonable results regarding geometry relaxation, vibrational spectrum and relaxation dynamics were obtained (ref. 3 in main text). We employ a similar procedure as in $[Dy(Cp^{ttt})_2]^+$, namely, the functional was PBEPBE, complemented with the GD3BJ empirical dispersion. Basis sets: cc-pVDZ for carbon and hydrogen atoms, and the effective core potential Stuttgart-RSC-1997-ECP for the uranium atom.

# Generation of distorted geometries. Determination of the CFP second derivatives at the relaxed geometry.

From the vibrational spectrum calculation (see output uranocenium-r.log), we get a $3N$-dimensional Cartesian displacement vector $\vec{w_j}$ for each vibrational mode $j = 1, ..., R$, where $N$ is the number of vibrating atoms and $R$ is the number of vibrational modes. Given the $3N$-dimensional Cartesian vector $\vec{v_{eq}}$ that contains the atom coordinates of the relaxed geometry, each distorted geometry $d$, represented by the $3N$-dimensional Cartesian vector $\vec{v_j^d}$, is generated as $\vec{v_j^d} = \vec{v_{eq}} + Q_j^d \vec{w_j}$, where $Q_j^d$ is a given real value of the distortion coordinate $Q_j$ of the vibrational mode $j$. Note that when $Q_j^d = 0$, the relaxed geometry is recovered $\vec{v_j^d} = \vec{v_{eq}}$. We are going to explain now how we choose the set of values $\{Q_j^d\}_d$ for a given vibrational mode $j$. First, we calculate the minimum value $Q_j^{\min}$, i.e., the minimum displacement of the given vibrational mode $j$. The key idea is to produce a significant distortion respect to the relaxed geometry. In other words, the distortion must produce on each component of each vibrating atom position a change at least equal to the experimental crystallographic error of this component. Since each displacement vector is normalized, we have $Q_j^{\min} = \left\| \vec{v_j^{\min}} - \vec{v_{eq}} \right\|$, i.e., $Q_j^{\min} = \sqrt{\sum_{i=1}^{N} \sum_{\alpha=x,y,z} \left( \alpha_j^{i,\min} - \alpha_{eq}^i \right)^2}$, where $\left( x_{eq}^i, y_{eq}^i, z_{eq}^i \right)$ is the equilibrium position of the atom $i$, and $\left( x_j^{i,\min}, y_j^{i,\min}, z_j^{i,\min} \right)$ is the minimum distorted position of the atom $i$ under the vibrational mode $j$. As said, each $\alpha_j^{i,\min} - \alpha_{eq}^i$ must equal $\delta\alpha^i$, which is the experimental crystallographic error in the $\alpha$ component of the atom $i$. Thus, we propose that the minimum displacement $Q_j^{\min}$ for the given vibrational mode $j$ be:

$Q_j^{\min} = \sqrt{\sum_{i=1}^{N} \sum_{\alpha=x,y,z} \left( \delta\alpha^i \right)^2}$. Standard X-ray crystallographic techniques do not usually detect hydrogen atoms because of their low electron density, and the relevant post-processing software does not attribute any crystallographic error to the hydrogen atoms in the experimental structure. Thus, we decide not to include any hydrogen atom in the expression of $Q_j^{\min}$. Sometimes, there exist components in the displacement vectors $\vec{w_j}$ that are zero or close enough to zero. This means that some components of some atoms do not change or hardly change from their equilibrium values when the vibrational mode $j$ is working. We also decide to exclude from the $Q_j^{\min}$ expression those atom components which do not change or hardly change. To decide which atom components must be excluded, we use a threshold in the corresponding component of the displacement vector. If the absolute value of the corresponding component in the displacement vector is not greater than the threshold, then we set $\delta\alpha^i = 0$, see below.

First, to determine the minimum displacements $Q_j^{\min}$ we include the code prefactor-v2.f90 in the folder Home-Made-Codes, along with the input prefactor-v2-input and the output prefactor-v2-output. Inside the code, one finds the variables nmod (number of vibrational modes), nato

(number of atoms, namely, all vibrating atoms but the hydrogen atoms), and the threshold th to decide whether a given component of a given displacement vector is small enough or not. In the input prefactor-v2-input, one first needs to introduce the three lattice parameters (in Angstroms) $a$, $b$, $c$ of the experimental structure. Since we lack an experimental structure of [U(Cp$^{ttt}$)$_2$]$^+$, we used the lattice parameters of [Dy(Cp$^{ttt}$)$_2$]$^+$. Note that the protocol to choose the $Q_j^d$ values for each vibrational mode $j$ is not unique, we only need some distorted geometries around the relaxed geometry in order to calculate the CFPs second derivatives respect to the distortion coordinate for each vibrational mode $j$. Thus, it is not strictly necessary to have some of the real parameters of the system under study. We need the lattice parameters (in Angstroms) because in prefactor-v2-input we introduce the crystallographic errors in fractional coordinates, and $Q_j^{\min}$ must be calculated in Angstroms. Whenever the crystallographic errors are introduced in orthogonal coordinates in Angstroms, one will set the three lattice parameters to be 1.0. After the lattice parameters in prefactor-v2-input, we introduce the crystallographic errors in fractional coordinates as said. The first row is for the metal, and the following rows are for the remaining atoms, in this case, carbon atoms. These crystallographic errors are from the [Dy(Cp$^{ttt}$)$_2$]$^+$ experimental structure. The error in the x-component is more or less the same for all the carbon atoms. Thus, we used the average value for the x-component of the error in each carbon atom. We analogously proceed for the y and z components of the error. To end up, after the crystallographic errors, the displacement vectors are introduced as directly provided by the output of Gaussian09 (keep the blank lines between lattice parameters, crystallographic errors and displacement vectors). From these displacement vectors, we must to remove those rows corresponding to the hydrogen atoms whenever they exist. The order both in the crystallographic error rows and in the displacement vectors rows has to be the same. For example, since the first row in each one of our displacement vectors corresponds to the metal atom, the first row of the crystallographic errors table must also correspond to the metal atom. Since we gave the same crystallographic error to each carbon atom and since all remaining rows of the displacement vectors (once those rows corresponding to hydrogen atoms have been removed) correspond to carbon atoms, there is no need to care about the order of the remaining rows in the crystallographic errors table. The lapack library is not required. To run the code, we use the compiler gfortran and the command line: gfortran –o aa prefactor-v2.f90 (aa is the executable name). In the output prefactor-v2-output, each minimum displacement $Q_j^{\min}$ is given after printing each displacement vector as "minimum displacement (Angstroms)".

Second, we calculate how many distorted geometries we need for each vibrational mode $j$. For that purpose, we include the excel file Number-Distorted-Geom in the folder Home-Made-Codes. In this file, the input consists of the harmonic frequencies ν and the force constants k as directly provided by the Gaussian09 output uranocenium-r.log. If the values of ν and k are changed, the values of l.c.(n=0) and l.c.(n=1) will be automatically changed. The variables l.c.(n=0) and l.c.(n=1) are the classical limits (Angstroms) in the distortion coordinate corresponding to the ground and first excited harmonic vibrational levels of a given vibrational mode $j$. Let us recall that the minimum displacement $Q_j^{\min}$ (Angstroms) for each vibrational mode $j$ is provided by the code prefactor-v2.f90. We take the maximum displacement (Angstroms) of the distortion coordinate of a given vibrational mode $j$ as the value $Q_j^{\min} \cdot s_j$, where $s_j$ is the natural number such that $Q_j^{\min} \cdot s_j$ is the smallest real number above l.c. (n=0). Thus, the number of distorted geometries for the given vibrational mode $j$ is $2 \cdot s_j$.

To end up, we need to generate the distorted geometries for each vibrational mode. For that, we include the code geom_dist.f90 in the folder Home-Made-Codes. In the input geom_dist_input, we first provide the minimum displacement $Q_j^{\min}$ (Angstroms) and the natural number $s_j$ for each vibrational mode $j$. These pairs -each one of them corresponds to a given vibrational mode $j$- have to be ordered with increasing frequency. Then (keep the blank line), we write the relaxed geometry in Cartesian coordinates (Angstroms) along with the chemical symbol of each atom. And then (keep again the blank lines), we write the displacement vectors as directly provided by Gaussian09. These displacement vectors have to also appear with increasing frequency. Let us recall that each row of each displacement vector corresponds to a vibrating atom. Thus, the order of the rows in each displacement vector has to be the same as that of the relaxed geometry. For example, the first row in our relaxed geometry is for uranium. Thus, the first row of all displacement vectors must also correspond to uranium. We suggest to introduce both the relaxed geometry and the displacement vectors in this input as directly provided by the Gaussian09 output. The code geom_dist.f90 provides the distorted geometries of each vibrational mode with the format required by the input simpre.dat in the package SIMPRE1.2 (ref. 12 in main text). Thus, the first row both in the relaxed geometry and in each displacement vector must always correspond to the metal. Moreover, all the metal-coordinating ligand atoms must be grouped together and must appear both in the relaxed geometry and in each displacement vector right after the first row (which corresponds to the metal). We suggest to build the Gaussian09 input such that the input geometry have the metal in the first row and the metal-coordinating ligand atoms grouped together in the following rows. Inside the code geom_dist.f90, there are some parameters: nmod is the number of vibrational modes, nato is the number of all vibrating atoms (including now also the hydrogen atoms whenever they are present), iesf decides whether the relaxed geometry and the distorted geometries are printed in Cartesian (Angstroms) or spherical coordinates (Angstroms, degrees, degrees), imax is the number of metal-coordinating ligand atoms plus one (i.e., this number exactly includes the metal and the metal-coordinating ligand atoms, only the rows from 2 to imax of the relaxed geometry and of the distorted geometries will be printed), rres is the radial displacement (Angstroms) to approach each metal-coordinating ligand atom to the metal ion (see REC model below), ceff is the effective charge to apply to each metal-coordinating ligand atom (see REC model below). The angles "the" and "phi" are used to rotate the relaxed geometry and the distorted geometries if desired. First, a clock-wise rotation of angle "the" (degrees) is performed around the Y=(0,1,0) axis. Then, an anticlock-wise rotation of angle "phi" (degrees) is performed around the Z=(0,0,1) axis. If this rotation is not desired, just set the=0.0d0 and phi=0.0d0. The rotation we use (the=84.9d0, phi=26.9d0 in degrees) is such that the two carbon rings that coordinate the metal become parallel to the XY plane. This rotation option only works when iesf=1. In the output geom_dist_output, one finds the relaxed geometry and the distorted geometries for each vibrational mode and for each value of the distortion coordinate. The rows containing the atom positions are ready to be used in the input simpre.dat in the package SIMPRE1.2 (ref. 12 in main text). To run geom_dist.f90, the library lapack is not required. We use the compiler gfortran and the command line: gfortran –o aa geom_dist.f90 (aa is the executable name).

All in all, for each mode $j$, several distorted geometries $\left\{\vec{v}_d^j\right\}_d$ are generated around the relaxed geometry $\vec{v}_{eq}$ by following the corresponding displacement vector $\vec{w}_j$. Then, we run a SIMPRE calculation at each distorted geometry $\vec{v}_j^d$ to determine the set of CFPs $A_k^q \left\langle r^k \right\rangle$ in cm-

[1] by following the same procedure employed to determine $\left\{\left(A_k^q \langle r^k \rangle\right)_{eq}\right\}_{k,q}$. Namely, to apply the same radial distance variations and charge magnitudes to the metal-coordinating atoms in these distorted geometries. Thus, for each mode $j$ and each CFP we have a set of pairs $\left\{\left(\left(A_k^q \langle r^k \rangle\right)_j^d, Q_j^d\right)\right\}_d$. In the folder CFPs-vs-Dist-Coord, one finds, for each vibrational mode, these CFPs (cm$^{-1}$) at each value of the distortion coordinate (Angstrom). The files in this folder are ready to plot each CFP evolution against the corresponding evolution of the distortion coordinate, for example, by using the Igor package. By fitting each plot "$\left(A_k^q \langle r^k \rangle\right)_j^d$ vs $Q_j^d$" to a second order polynomial and evaluating its second derivative at $Q_j = 0$ (let us recall that the second derivatives in Eq. 1 are evaluated at the relaxed geometry, which corresponds to $Q_j = 0$ for all $j$) we access $\left(\partial^2 A_k^q \langle r^k \rangle / \partial Q_j^2\right)_{eq}$. The values of these second derivatives for each CFP and for each vibrational mode are provided in the input frrmcfpd2.inp (which can be found in the folder Home-Made-Codes). In general, to fit each plot "CFP vs $Q_j$", we use the smallest degree polynomial that provides the best visual and most reasonable fitting (see SI of ref. 9 in main text).

## Transition rates

The system vibrations that perturb the equilibrium electronic structure (corresponding to the relaxed geometry) are considered as harmonic. To determine the probability per unit time (i.e., the transition rate) of driving a transition from a state of the system characterized by the ket $|E_i, n_j\rangle$, where $|E_i\rangle$ is a given initial eigenstate with energy $E_i$ and the quantum number $n_j$ describes the eigenstate of a given 1D harmonic vibrational mode $j$, to another state $|E_f, n_j \pm 1\rangle$ either by emitting or by absorbing a phonon, where $E_f$ is the energy of a given final eigenstate, is common to proceed by employing the so-called Fermi Golden Rule. This rule is usually prepared to incorporate a given expression of the phonon density of states. For example, the most employed phonon density of states is that of the Debye model, where this density is proportional to the square of the phonon frequency. Then, one integrates the transition rate over this phonon frequency up to the Debye cut-off, and the resultant expression depends on some parameters such as the crystal longitudinal and transverse sound velocities besides the Debye temperature. Let us recall that the current main goal is the development of fully ab-initio methodologies, in particular, by not assuming any specific form in the phonon density of states. This means being able to incorporate the vibrational spectrum as provided by a first-principles calculation. In solid state systems, vibrational energies are close enough so that it is considered they form an energy continuum. That is why the phonon frequency in the Debye model appears as a continuous variable -not discrete- which is subsequently integrated over a given real interval. On the contrary, a first-principles software will always provide a finite number of vibrations (each one with its harmonic frequency, reduced mass, force constant and displacement vector). These vibrations are the result of diagonalizing the so-called force matrix, which has always a finite size because computers only deal with finite quantities. To incorporate this finite set of vibrations into the transition rates, we need to replace the standard integral of the phonon frequency over a real interval by a summation over this given set of vibrations. Thus, the transition rates we show below are the result of adapting the standard Fermi Golden Rule, where the integral over the phonon frequency have been substituted by a summation over all vibrational modes (ref. 3 in main text). Indeed, the spin-vibration coupling is calculated up to second order in perturbation theory. Thus, there no exist crossed interactions among different vibrational modes and the transition rate expressions are just a summation over independent vibrations (ref. 9 in main text). These transition rates have been derived under the so-called Born-Oppenheimer approximation, which assumes that the electronic and nuclear dynamics are uncoupled and it results in non-adiabatic electronic transitions.

Orbach transition rates:

This relaxation process is a finite sequence of direct transitions $|i\rangle \to |f\rangle$ where each one of them is driven by only one resonant phonon with the energy difference $|E_f - E_i|$. The process starts in an initial eigenstate with unity population. The spin is excited to higher intermediate eigenstates in the potential barrier through phonon absorption. Once the barrier has been crossed (either by overcoming the highest eigenstate or by tunneling), this is followed by a cascade of de-excitations until reaching a final eigenstate through phonon emission.

Phonon absorption: $\gamma_{fi} = \dfrac{2\pi}{\hbar} \sum_{j=1}^{R} \left[ |\langle i|\hat{H}_j|f\rangle|^2 |\langle n_j - 1|\varepsilon_j|n_j\rangle|^2 \rho_j(|E_i - E_f|) \right]$  Eq. S1

Phonon emission: $\gamma_{if} = \dfrac{2\pi}{\hbar} \sum_{j=1}^{R} \left[ |\langle i|\hat{H}_j|f\rangle|^2 |\langle n_j + 1|\varepsilon_j|n_j\rangle|^2 \rho_j(|E_i - E_f|) \right]$  Eq. S2

$R$ is the number of vibrational modes.

Second-order Raman transition rates:

The transition from $|i\rangle$ to $|f\rangle$ is not direct but driven through an intermediate eigenstate $|c\rangle$, and involves two resonant phonons with the energy differences $|E_c - E_i|$ and $|E_f - E_c|$. The first phonon $j$ mixes $|i\rangle$ with $|c\rangle$, while the second one $l$ mixes $|c\rangle$ with $|f\rangle$. Now, the case $E_i = E_f$ will have a certain transition rate whose value is not necessarily zero. Given $|i\rangle$ and $|f\rangle$, we include in the transition rate expression all intermediate eigenstates $|c\rangle$ but the ones with an energy $E_c$ equal to either $E_i$ or $E_f$ (ref. 3 in main text). Thus, given $|i\rangle$ and $|f\rangle$, for each $|c\rangle$ only one of the following four options regarding the order in the energies is possible: (i) $E_i < E_c > E_f$, (ii) $E_i > E_c < E_f$, (iii) $E_i > E_c > E_f$, (iv) $E_i < E_c < E_f$. The transition rate expression is as follows:

$$\gamma = \dfrac{2\pi}{\hbar} \sum_{\substack{c=1 \\ E_c \neq E_i, E_f}}^{2J+1} \sum_{j=1}^{R} \sum_{l=1}^{R} \left[ \dfrac{|\langle c|\hat{H}_j|i\rangle|^2 |\langle f|\hat{H}_l|c\rangle|^2}{|E_c - E_i||E_f - E_c|} \Theta(c,j)\Theta(c,l) \rho_j(|E_c - E_i|) \rho_l(|E_f - E_c|) \right]$$

Eq. S3

In case of (i), the phonon $j$ is absorbed and the phonon $l$ is emitted; thus $\Theta(c,j) = |\langle n_j - 1 | \varepsilon_j | n_j \rangle|^2$ and $\Theta(c,l) = |\langle n_l + 1 | \varepsilon_l | n_l \rangle|^2$. In case of (ii), the phonon $j$ is emitted and the phonon $l$ is absorbed; thus $\Theta(c,j) = |\langle n_j + 1 | \varepsilon_j | n_j \rangle|^2$ and $\Theta(c,l) = |\langle n_l - 1 | \varepsilon_l | n_l \rangle|^2$. In case of (iii), the phonon $j$ is emitted and the phonon $l$ is emitted; thus $\Theta(c,j) = |\langle n_j + 1 | \varepsilon_j | n_j \rangle|^2$ and $\Theta(c,l) = |\langle n_l + 1 | \varepsilon_l | n_l \rangle|^2$. In case of (iv), the phonon $j$ is absorbed and the phonon $l$ is absorbed; thus $\Theta(c,j) = |\langle n_j - 1 | \varepsilon_j | n_j \rangle|^2$ and $\Theta(c,l) = |\langle n_l - 1 | \varepsilon_l | n_l \rangle|^2$.

Matrix elements of the strain tensor:

$$|\langle n_j - 1 | \varepsilon_j | n_j \rangle|^2 = \frac{1}{e^{\hbar\omega_j/k_B T} - 1} \quad |\langle n_j + 1 | \varepsilon_j | n_j \rangle|^2 = \frac{1}{1 - e^{-\hbar\omega_j/k_B T}} \quad \text{Eq. S4}$$

These matrix elements describe the strain suffered by the lattice, which is encoded in the so-called strain tensor $\varepsilon_j$, when the vibrational mode $j$ absorbs or emits a phonon, respectively. The vibration bath is considered to be thermalized, i.e., its dynamics is much faster than that of the magnetic relaxation. Thus, these matrix elements are proportional to the Bose-Einstein statistics of the given vibrational mode $j$, and depend only on temperature (ref. 21 and 23 in main text).[2] Note that when temperature $T \rightarrow 0$ the left matrix element in Eq. S4 vanishes, but not the right term which tends to 1. This means that some transition rates do not necessarily vanish as $T \rightarrow 0$, and thus the spin is expected to relax even at very low temperature.

Distribution of phonon energies:

$$\rho_j(\hbar\omega) = \frac{1}{\sigma\sqrt{2\pi}} \exp\left(-\frac{1}{2}\left(\frac{\hbar\omega - \hbar\omega_j}{\sigma}\right)^2\right), \quad j = 1, ..., R \quad \text{Eq. S5}$$

There is another modification implemented in the above transition rates. The original expressions contain the Dirac delta function $\delta(\hbar\omega - \hbar\omega_j)$, where $\hbar\omega = |E_f - E_i|$ is the energy difference between the final and the initial eigenstates. The conservation of energy implies that $\hbar\omega$ must equal the phonon energy $\hbar\omega_j$ of a given vibrational mode $j$. Otherwise, both $\delta(\hbar\omega - \hbar\omega_j)$ and the corresponding transition rate vanish, i.e., there is no spin transition. As said above, first-principles packages provide a discrete vibrational spectrum. Thus, it is quite unlikely to find a vibrational mode whose phonon energy exactly matches a given energy difference $|E_f - E_i|$, and hence one would not observe any spin relaxation. To solve this issue (ref. 3 in main text), the Dirac delta function is replaced by a Gaussian convoluted spectrum around the phonon energy $\hbar\omega_j$ of the given vibrational mode $j$. In other words, we let the phonon energy have an uncertainty width around its value $\hbar\omega_j$. This width is determined by the standard deviation parameter $\sigma$, and can be estimated by inspecting the experimental IR and Raman vibrational spectra (the full-width-half-maximum linewidth is twice as much as $\sigma$) (ref. 3 in main text). This parameter has to be estimated carefully, since a too small value makes the Gaussian convoluted spectrum become too much similar to a delta function, and no relaxation is observed. On the contrary, a too large value means a continuously flat vibrational spectrum, which is not observed for molecular systems. Our case study just aims to demonstrate the methodology that we are proposing, and we have no experimental vibrational spectra of [U(Cp$^{ttt}$)$_2$]$^+$. Thus, we decided to employ the same value as in [Dy(Cp$^{ttt}$)$_2$]$^+$, which is $\sigma \sim 10 cm^{-1}$.

## Resolution of the master equation

It is important to stress that the master equation in Eq. 2 is not invariant by time reversal. Thus, it is only valid for large enough times when irreversibility in the macroscopic system has been established. Irreversibility is reached above the time scale in which the relevant system-environment collisions occur. Since spin-vibration interactions are much faster than relaxation in magnetic molecules, we can safely assume the attainment of this macroscopic irreversibility (ref. 21 in main text). A more complete description of relaxation would need to use density matrix formalism, since the employed picture consists in a spin population flowing among the several eigenstates, which thus disregards any coherent superposition of them.

Orbach process:

We explain now how to solve the master equation in Eq. 2 (ref. 3, 21, 22 in main text). For that, we make use of the Orbach transition rates in Eq. S1 and Eq. S2. First, we need to build the so-called master matrix $\Gamma$ from the transition rates, whose size is $(2J+1)\times(2J+1)$ ($J$ is the ground electron spin quantum number of the magnetic metal ion). As explained in the main text, after diagonalizing the equilibrium crystal field Hamiltonian we obtain the lowest $2J+1$ eigenstates, which are truncated to the $|m_J\rangle$ components of the ground $J$ multiplet and subsequently renormalized in the case of studying a U$^{3+}$-based molecular magnet. Then, these eigenstates $|e\rangle$ must be ordered. In our case, the order we consider is the one provided in the input rates.inp for the code rates.f (see section "Fortran code to evaluate magnetic relaxation dynamics"). In this input, we write the eigenstates row-wise; the row $1$ is for the first eigenstate $|e_1\rangle$ and the row $2J+1$ is for the last one $|e_{2J+1}\rangle$. The master matrix $\Gamma$ is:

| $\Gamma$ | $|e_1\rangle$ | $|e_2\rangle$ | ... | $|e_{2J}\rangle$ | $|e_{2J+1}\rangle$ |
|---|---|---|---|---|---|
| $|e_1\rangle$ | | | | | |
| $|e_2\rangle$ | | | | | |
| ... | | | | | |
| $|e_{2J}\rangle$ | | | | | |
| $|e_{2J+1}\rangle$ | | | | | |

First, we fill in the off-diagonal elements. Given initial and final eigenstates $|i\rangle$ and $|f\rangle$, the positions of $|i\rangle$ and $|f\rangle$ in $\Gamma$ determine the column and the row of the off-diagonal element to fill in, resp. If $E_f > E_i$, we use the transition rate $\gamma_{fi}$ (phonon absorption). If $E_f < E_i$, we use the transition rate $\gamma_{if}$ (phonon emission). If $E_f = E_i$, we use $\gamma = 0$ (ref. 3 in main text, we are not including quantum tunneling of magnetization due to the action of a spin bath). Now, if the off-diagonal element $(|f\rangle, |i\rangle)$ has been filled with $\gamma_{fi}$, the symmetric off-diagonal element $(|i\rangle, |f\rangle)$ is filled with $\gamma_{if}$. If the off-diagonal element $(|f\rangle, |i\rangle)$ has been filled with $\gamma_{if}$, the symmetric off-diagonal element $(|i\rangle, |f\rangle)$ is filled with $\gamma_{fi}$. If the off-diagonal element $(|f\rangle, |i\rangle)$ has been filled with $\gamma = 0$, the symmetric off-diagonal element $(|i\rangle, |f\rangle)$ is also filled with $\gamma = 0$. Each diagonal element in a given position $(|e\rangle, |e\rangle)$ is the negative summation of the off-diagonal elements in the given $|e\rangle$ column. By numerical diagonalization, $2J+1$ real eigenvectors $(\varphi_{e_1,k}, ..., \varphi_{e_{2J+1},k})$ and eigenvalues $-1/t_k$ are obtained, where $t_k$ are the positive $2J+1$ relaxation times of the system ($-1/t_k$ are thus the $2J+1$ relaxation rates of the system). One of these rates $-1/t_{k_0}$ is always zero, and corresponds to the situation in which the system has reached the thermal equilibrium. From these formalism, an expression for the total magnetization $M(t)$ as a function of time is obtained, Eq. S6, which depends on the population $0 \leq p_e(t) \leq 1$ and on the magnetization $M_e$ of each eigenstate $|e\rangle$.

The expression of $p_e(t)$ is shown in Eq. S7, and is a finite sum proportional to the $2J+1$ exponential functions $e^{-t/t_k}$. The coefficients $\{\lambda_k\}_{k=1}^{2J+1}$ are obtained by solving the linear equation system in Eq. S7 after setting $t = 0$ and introducing the $2J+1$ initial populations $0 \leq p_e(t=0) \leq 1$. The magnetizations $M_e$ are calculated as the derivatives $M_e = -\frac{\partial}{\partial B}\langle e|\hat{H}_{ZE}|e\rangle$ evaluated at $B = 0$, where $B$ is a static magnetic field and $\hat{H}_{ZE}$ is the Zeeman Hamiltonian.[3] After applying the Hellmann-Feynman theorem, $M_e$ can be rewritten as $M_e = -\langle e|\frac{\partial \hat{H}_{ZE}}{\partial B}|e\rangle$. We consider that the given magnetic molecule shows an axial anisotropy, with an axis either easy or hard that defines the Z axis. The magnetic field is applied in this Z direction, and thus the Zeeman Hamiltonian is $\hat{H}_{ZE} = \mu_B g B \hat{J}_z$, where $\mu_B$ is the Bohr magneton and $g$ is the free-ion Landé factor. Hence, $M_e = -\mu_B g \langle e|\hat{J}_z|e\rangle$. Let us write the eigenstate $|e\rangle$ in the basis set of the ground $J$ multiplet as $|e\rangle = \sum_{j=1}^{2J+1} c_e(j)|j-J-1\rangle$, where $c_e(j)$ are complex coefficients such that $|e\rangle$ is normalized. After some algebra, it is easy to obtain the expression $\langle e|\hat{J}_z|e\rangle = \sum_{j=1}^{2J+1} |c_e(j)|^2 (j-J-1)$ for the expectation value of the z component of the electron spin operator $\hat{J}$.

$$M(t) = \sum_{e=1}^{2J+1} p_e(t) M_e \quad \text{Eq. S6}$$

$$p_e(t) = \sum_{k=1}^{2J+1} \lambda_k \varphi_{e,k} e^{-t/t_k} \quad \text{Eq. S7}$$

In practice, $2J-1$ of the exponential functions in Eq. S7 vanish quickly (there are $2J-1$ relaxation rates that are extremely fast), and $M(t)$ behaves as a single decaying exponential (corresponding to the only one slow relaxation rate) plus a constant derived from $-1/t_{k_0} = 0$. This fact is due to the double-well anisotropy. Indeed, each one of these $2J-1$ fast relaxation rates describes spin dynamics inside a given side of the potential barrier. On the contrary, the spin jump over the barrier takes a much longer time and is accounted for by the slowest relaxation rate.

Since $g$ is independent of the script $e$, we redefine the magnetization $M_e$ as $\tilde{M}_e = M_e / \mu_B g$. Thus, $\tilde{M}_e = -\sum_{j=1}^{2J+1} |c_e(j)|^2 (j - J - 1)$. We can also redefine the total magnetization $M(t)$ as $\tilde{M}(t) = M(t)/\mu_B g$. Thus, $\tilde{M}(t) = -\sum_{e=1}^{2J+1} \left( p_e(t) \sum_{j=1}^{2J+1} |c_e(j)|^2 (j - J - 1) \right)$. By fitting the plot "$\tilde{M}(t)$ vs $t$" to an exponential function $f(t) = a + be^{-t/\tau}$, the overall Orbach-mediated magnetic relaxation time $\tau$ is extracted, which is dominated by the slowest relaxation rate $-1/t_k$ at the working temperature as explained.

Second-order Raman process:

Each off-diagonal element (column $|i\rangle$, row $|f\rangle$) of the master matrix is filled by calculating the transition rate given in Eq. S3. The diagonal elements are also built by adding the elements of the corresponding column and changing the sign of the summation. The master matrix is diagonalized, which allows us reaching the total magnetization whose time decay is fitted to a single exponential function to extract the Raman-mediated magnetic relaxation time at the working temperature.

In the following table, we show the second-order Raman relaxation time calculated at different temperatures.

Table S1. Thermal evolution of the second-order Raman relaxation time.

| Temperature (K) | Second-order Raman Relaxation Time (s) |
|---|---|
| 11 | 3,4·10$^7$ |
| 12 | 5,1·10$^6$ |
| 13 | 1,1·10$^6$ |
| 14 | 2,9·10$^5$ |
| 15 | 9,3·10$^4$ |
| 20 | 1,7·10$^3$ |
| 25 | 1,5·10$^2$ |
| 30 | $2{,}9 \cdot 10^1$ |
| 40 | 3,6 |
| 50 | 1,0 |

The reason to calculate the Orbach and second-order Raman relaxation times up to 50 K is because at this temperature the Orbach relaxation time (Fig. 3) is already reaching the lowest value for a relaxation time that is experimentally observed, which is around $10^{-5}$ s. The reason to calculate the relaxation times above 6 K in case of the Orbach process (Fig. 3) and above 11 K in case of the second-order Raman process is because below these temperatures the smallest relaxation rates (eigenvalues of the corresponding master matrix) cannot be distinguished from the computational numerical noise. For example, below these temperatures one finds that the two smallest relaxation rates are both positive. It is important to note that only one relaxation rate $-1/t_{k_0}$ may be positive as much, and this rate in fact should be strictly zero (see above). Nevertheless, because of the computational numerical noise, it may have a rather non-zero value. That is why we introduce the thresholds eot (in case of the Orbach master matrix) and ert (in case of the second-order Raman master matrix), which have to be given a value such that they are above the magnitude of the smallest relaxation rate and below the magnitude of the second smallest relaxation rate (see section "Fortran code to evaluate magnetic relaxation dynamics"). The remaining $2J$ relaxation rates should be all of them always real and negative. Below these temperatures, we have also found (i) imaginary relaxation rates (although they are rather small and comparable to the computational numerical noise), (ii) the two smallest relaxation rates which are too similar to be distinguished and to decide which one should be strictly zero, and (iii) the smallest non-zero relaxation rates too small that makes the magnetization not to decay with time (which is interpreted as an infinite relaxation time).

## Determination of relaxation pathways and identification of vibrations promoting relaxation

To get further insight into the spin dynamics, a proper re-cast of the Orbach master matrix $\Gamma$ provides the different relaxation pathways at the working temperature. In this master matrix, the ket that determines a given column is the initial eigenstate $|i\rangle$ of a given direct transition, while the ket of a given row is the final eigenstate $|f\rangle$ of the same direct transition. At time $t=0$, we place all the initial population in the [U(Cp$^{ttt}$)$_2$]$^+$ eigenstate with $\langle \hat{J}_z \rangle = -3.99$, see Fig. 4. Note that in Fig. 4 the eigenstates are not ordered from left to right by increasing $\langle \hat{J}_z \rangle$ value. Nevertheless, let us consider the appearance order from left to right of each one of these eigenstates in this figure. Thus, the eigenstate with $\langle \hat{J}_z \rangle = -3.99$ is first, the one with $\langle \hat{J}_z \rangle = -1.43$ is second, and so on until the final eigenstate with $\langle \hat{J}_z \rangle = +3.99$. We only consider those relaxation pathways whose direct transitions are always to a final eigenstate $|f\rangle$ which is to the right of the given initial eigenstate $|i\rangle$ in Fig. 4 (ref. 3 in main text). For example, from the eigenstate with $\langle \hat{J}_z \rangle = -3.99$ the direct transitions can be to any other eigenstate but the one with $\langle \hat{J}_z \rangle = +3.99$ (let us recall that in this approach we are not considering direct transitions between degenerate eigenstates), while from the eigenstate with $\langle \hat{J}_z \rangle = +0.16$ the direct transitions can only be to the one with $\langle \hat{J}_z \rangle = +1.43$ or to the one with $\langle \hat{J}_z \rangle = +3.99$. Sometimes (ref. 3 in main text), these equilibrium eigenstates may be pure in the sense that their $\hat{J}_z$ expectation values coincide with the several $m_J = -J,...,+J$ values. In this case, they would be ordered from left to right by increasing $\langle \hat{J}_z \rangle$ value, and to re-cast the Orbach master matrix only those direct transitions that increase the $\langle \hat{J}_z \rangle$ value (but the ones between degenerate eigenstates) would be considered, where the $\langle \hat{J}_z \rangle = -J$ eigenstate would be the one with unit population at $t = 0$. In our case, to do the re-cast, in the Orbach master matrix we only need to keep those entries that correspond to a direct transition between non-degenerate eigenstates, where the final eigenstate $|f\rangle$ is found to the right of the initial eigenstate $|i\rangle$ in Fig. 4. The remaining entries, including the ones in the main diagonal, are now filled with a zero (transitions from a given eigenstate to the same eigenstate are excluded). In the input rates.inp of the code rates.f, right after writing the initial populations at time $t = 0$, one writes a table whose size is the same as that of the Orbach master matrix (see section "Fortran code to evaluate magnetic relaxation dynamics"). We write "1" at those positions that correspond to those direct transitions that are kept; at the remaining positions we write "0". The re-casted Orbach master matrix appears in the output rates.out, in the section "Orbach relaxation pathway". Once the Orbach master matrix is re-casted, it needs now to be normalized in terms of percentages. The normalized re-casted Orbach master matrix appears in rates.out right after the re-casted Orbach master matrix. For that, right after the re-casting table just mentioned, we first write in rates.inp the total number of possible direct transitions between <u>consecutive</u> eigenstates from left to right in Fig. 4. Since there are 10 eigenstates, this total number is 9. Then, after this number we write the order of the equilibrium eigenstates as they appear in Fig.4 from left to right. The rightmost eigenstate $\langle \hat{J}_z \rangle = +3.99$ is excluded because there are no more eigenstates to the right, hence, there are no possible direct transitions from this eigenstate to keep. This order depends on the order with which the eigenstates were introduced at the beginning of rates.inp. In this input, we wrote the eigenstates in this order: 1 $\langle \hat{J}_z \rangle = -3.99$, 2 $\langle \hat{J}_z \rangle = +3.99$, 3 $\langle \hat{J}_z \rangle = -1.43$, 4 $\langle \hat{J}_z \rangle = +1.43$, 5 $\langle \hat{J}_z \rangle = -0.16$, 6 $\langle \hat{J}_z \rangle = +0.16$, 7 $\langle \hat{J}_z \rangle = -1.02$, 8 $\langle \hat{J}_z \rangle = +1.02$, 9 $\langle \hat{J}_z \rangle = +1.98$, 10 $\langle \hat{J}_z \rangle = -1.98$. The order we need to write in rates.inp is: $\langle \hat{J}_z \rangle = -3.99$ -> $\langle \hat{J}_z \rangle = -1.43$ -> $\langle \hat{J}_z \rangle = -0.16$ -> $\langle \hat{J}_z \rangle = -1.02$ -> $\langle \hat{J}_z \rangle = -1.98$ -> $\langle \hat{J}_z \rangle = +1.98$ -> $\langle \hat{J}_z \rangle = +1.02$ -> $\langle \hat{J}_z \rangle = +0.16$ -> $\langle \hat{J}_z \rangle = +1.43$ -> $\langle \hat{J}_z \rangle = +3.99$. Thus, we write in rates.inp the sequence 1 3 5 7 10 9 8 6 4 (note that the number 2 which corresponds to the $\langle \hat{J}_z \rangle = +3.99$ eigenstate is not present). The normalization in terms of percentages is based on the order of the sequence 1 3 5 7 10 9 8 6 4, and is as follows. These numbers in this sequence have now to be read as if they are referring to the corresponding columns of the re-casted Orbach master matrix (let us recall that the eigenstates that determine the several columns of this matrix are the initial eigenstates in the direct transitions). The first position in this sequence corresponds to the eigenstate with unity population at time $t=0$, and the populations of all direct transitions that departure from this eigenstate must amount to 100%. Thus, we sum all the elements in the column 1 of the re-casted Orbach master matrix, divide each one of these elements by this sum, and multiply each result by 100. After this process, each element of the first column in the normalized re-casted Orbach master matrix gives the population that flows from the eigenstate 1 ($\langle \hat{J}_z \rangle = -3.99$) to the rest of the eigenstates. In particular, the element (3,1) (row 3 and column 1) gives the population $P_{1 \to 3}$ that flows from the eigenstate 1 ($\langle \hat{J}_z \rangle = -3.99$) to the eigenstate 3 ($\langle \hat{J}_z \rangle = -1.43$). Now, the process is repeated in the column 3 of the re-casted Orbach master matrix. We add all the elements in the column 3, divide each element by this sum, and now multiply each result by $P_{1 \to 3}$. In this way, the sum of all the incoming populations to the eigenstate 3 ($\langle \hat{J}_z \rangle = -1.43$) equals the sum of all the outcoming populations from the same eigenstate, which is not necessarily 100. Now, each element of the third column in the normalized re-casted Orbach master matrix gives the population that flows from the eigenstate 3 ($\langle \hat{J}_z \rangle = -1.43$) to the rest of the eigenstates that are to the right in Fig. 4 (but the one with $\langle \hat{J}_z \rangle = +1.43$). The process is repeated again now with the column 5 of the re-casted Orbach master matrix, and finishes with the column 4 of this matrix. It is possible that the sum of the elements of a given column of the re-casted Orbach master matrix be below the computational numerical precision, which means this sum is as if it was zero. If this sum is zero from the point of view of the employed computer, each element of the given column can also be considered to be zero, since they are all always positive. Thus, it is not possible to divide these elements by the just mentioned sum. We introduce the threshold pot = 1.0d-16 in the code rates.f to decide when this sum has to be considered as zero. In this case, all the elements of the given column in the re-casted Orbach master matrix are set to be zero.

Once the re-casted Orbach master matrix has been normalized, each direct transition that compose each relaxation pathway can now be decomposed into the contributions from the several vibrational modes. Indeed, let us recall that each direct transition corresponds to an element of the re-casted Orbach master matrix. These elements are in fact transition rates, which are a sum over the several vibrational mode contributions, see Eq. S1 and Eq. S2. What we do in rates.f is to normalize each vibrational mode contribution in terms of percentages, i.e., we divide each contribution by the value of the given transition rate and then multiply the result by 100. If this percentage is above a threshold (between 0.0 and 100.0), then the vibrational mode number along with its contribution percentage is printed in the output rates.out (see section "Fortran code to evaluate magnetic relaxation dynamics"). This threshold is the last real number written in the input rates.inp. If a given transition rate in the re-casted Orbach master matrix is below the computational numerical precision, it can be considered to be zero. This means that each vibrational mode contribution to this transition rate can also be considered to be zero, since all of the contributions are positive. In this case, no vibrational modes corresponding to the given transition rate are shown in rates.out. To decide when a given transition rate has to be considered as zero, we use the threshold mct=1.0d-16 in the code rates.f. Once the most contributing vibrational modes are identified, we can now visually inspect how the molecule vibrates to check which atomic movements are involved. Then, chemical modifications in the molecular structure can be rationally proposed in order to remove these modes and suppress magnetic relaxation as much as possible, with the hope of improving the molecular magnet performance. In the outputs that are placed in the folder Orbach-rp-Raman, one can find those vibrational modes that contribute (>10%) to each direct transition of the Orbach-mediated relaxation pathways (see Fig. 4) at the several working temperatures.

Above T = 30 K, the thermally activated relaxation is at play. The spin population flows through excited doublets by absorption and emission of phonons (Fig. 4). We identify up to six vibrations involved in this relaxation mechanism. These are the calculated vibrations 16, 17, 18, 19, 20 ,21, with harmonic frequencies $v_{16}$ = 135.0115 cm$^{-1}$, $v_{17}$ = 136.8658 cm$^{-1}$, $v_{18}$ = 170.0364 cm$^{-1}$, $v_{19}$ = 172.5580 cm$^{-1}$, $v_{20}$ = 175.4401 cm$^{-1}$, $v_{21}$ = 175.7696 cm$^{-1}$ (see folder Animations), which closely match the gaps between the equilibrium ground and first excited doublets (159.3 cm$^{-1}$), and first and second excited doublets (171.7 cm$^{-1}$). The vibration 16 is a rocking-like deformation of the two Cp$^{ttt}$ rings: the two hydrogen atoms bounded to each Cp$^{ttt}$ ring moves towards and away from the U$^{3+}$ ion. As a side effect, there are also rigid movements of the terc-butyl substituents. This kind of vibration was also identified in a previous study on the molecular magnet [Dy(Cp$^{ttt}$)$_2$]$^+$ (calculated 64-67 vibrations of ref. 3 in main text) as the one promoting the first step in the most likely relaxation pathway from the ground doublet to the first excited doublet. It was proposed to substitute these two hydrogen atoms in the Cp$^{ttt}$ rings by bulkier substituents in order to block this vibration and, in fact, this substitution was carried out in a subsequent study (ref. 5 in main text). This modification worked since this vibration is no longer observed and, indeed, the blocking temperature is increased from 60 K to 80 K. The experimental effective barrier U$_{eff}$ is also increased from 1223 cm$^{-1}$ to 1541 cm$^{-1}$. The vibration 17 involves kind of rigid movements in the terc-butyl substituents. The modes 18 and 21 are symmetric and antisymmetric breathing vibrations: the two Cp$^{ttt}$ rings moves towards and moves away from the U$^{3+}$ ion at once and out of phase, respectively, and are also found in the recently reported molecular magent Dy-5* (vibrations 66, 67, 68 of ref. 5 in main text). This vibration could be suppressed by bounding these two Cp$^{ttt}$ rings, such as it happens in stapled bis-phthalocyanines. The vibrations 19 and 20 involve methyl rotations in the terc-butyl substituents. These rotations could be partially suppressed if one replaces the methyl groups –CH$_3$ by the heavier fluorinated analogs –CF$_3$.

## Fortran code to evaluate magnetic relaxation dynamics

In the folder Home-Made-Codes one finds the code rates.f, which is aimed to evaluate magnetic relaxation dynamics as we explain below, and the three inputs it needs, which are already prepared such as we used them: rates.inp, matrix.inp, frrmcfpd2.inp (matrix.inp is found in the subfolder Pert-Ham-Temp for each temperature, do not forget to rename this input before running rates.f). The code can generate up to three outputs: rates.orbach.out (see the folder Orbach-mag), rates.raman.out, and rates.out (see the folder Orbach-rp-Raman). To run rates.f, the library lapack is required. To compile, we use the compiler gfortran and the command line is: gfortran –o aa –llapack rates.f, where aa is the executable name. A useful information to run the codes is the numerical value of the ground electron spin quantum number $J$ of the selected magnetic metal ion: $J(Ce(III))=5/2$, $J(Pr(III))=4$, $J(Nd(III),U(III))=9/2$, $J(Pm(III))=4$, $J(Sm(III))=5/2$, $J(Tb(III))=6$, $J(Dy(III))=15/2$,

$J(Ho(III)) = 8$, $J(Er(III)) = 15/2$, $J(Tm(III)) = 6$, $J(Yb(III)) = 7/2$. We describe now these inputs, the code and the outputs:

frrmcfpd2.inp: this input contains the harmonic frequencies with increasing energy (cm$^{-1}$, as directly provided by the Gaussian09 output uranocenium-r1.log), reduced masses accordingly ordered with the harmonic frequencies (chemical atomic mass units, as directly provided by the Gaussian09 output uranocenium-r1.log), the second derivatives of the CFPs $A_k^q \langle r^k \rangle$ respect to the distortion coordinate of each vibrational mode evaluated at the relaxed geometry (cm$^{-1}$·Å$^{-2}$, the order of these derivatives is (2,0), (2,1), (2,-1), (2,2), (2,-2), (4,0), (4,1), (4,-1), (4,2), (4,-2), (4,3), (4,-3), (4,4), (4,-4), (6,0), (6,1), (6,-1), (6,2), (6,-2), (6,3), (6,-3), (6,4), (6,-4), (6,5), (6,-5), (6,6), (6,-6), where the parenthesis (k,q) represents the scripts k = 2, 4, 6, q = -k,..., +k), and the CFPs $\left( A_k^q \langle r^k \rangle \right)_{eq}$ determined at the relaxed geometry (cm$^{-1}$, given in the same order as that in each one of the second derivatives). Actually, rates.f only needs the harmonic frequencies from this input.

matrix.inp: this input contains the perturbing Hamiltonian $\hat{H}_j = \sum_{k=2,4,6} \sum_{q=-k}^{k} \Delta \left( A_k^q \langle r^k \rangle \right)_j (T) \eta_k \hat{O}_k^q$ of each vibrational mode $j$ at a given temperature $T$ (see Eq. 1 in main text), where $\eta_2 = \alpha$, $\eta_4 = \beta$, $\eta_6 = \gamma$ are the Stevens factors and $\hat{O}_k^q$ are the Stevens equivalent operators. These Hamiltonians in matrix.inp follow the harmonic frequencies order in frrmcfpd2.inp. Each perturbing Hamiltonian is a $(2J+1) \times (2J+1)$ complex matrix with the same units as the parameters $\Delta \left( A_k^q \langle r^k \rangle \right)_j (T)$ (here, in cm$^{-1}$), written in the ordered basis set $\{|-J\rangle,...,|+J\rangle\}$, where $J$ is the ground electron spin quantum number. The parameters $\Delta \left( A_k^q \langle r^k \rangle \right)_j (T)$ are determined by running the cfppert.f code (see the section "Fortran codes to calculate CFPs thermal evolution").

rates.inp:

The first row contains the lowest $2J+1$ energies (cm$^{-1}$) calculated at the relaxed geometry, i. e., those obtained by diagonalizing the equilibrium crystal field Hamiltonian $\hat{H}_{eq} = \sum_{k=2,4,6} \sum_{q=-k}^{k} \left( A_k^q \langle r^k \rangle \right)_{eq} \eta_k \hat{O}_k^q$, which is built with the CFPs $\left( A_k^q \langle r^k \rangle \right)_{eq}$ of the relaxed geometry. Let us recall that for U$^{+3}$-based molecular magnets the diagonalization is performed by using the CONDON package (ref. 20 in main text), where the Hamiltonian includes both the ground and excited $J$ multiplets and the CFPs must be introduced in Wybourne notation since this software uses a rather different implementation of the crystal field operators.

The following $2J+1$ rows contain the lowest $2J+1$ eigenstates calculated at the relaxed geometry, i. e., those obtained by diagonalizing the equilibrium crystal field Hamiltonian $\hat{H}_{eq}$ (see above). The coefficients are complex and such that each eigenstate is normalized. These $2J+1$ rows are accordingly ordered with the corresponding $2J+1$ energies. Let us recall that for U$^{3+}$-based molecular magnets (see above) the lowest $2J+1$ eigenstates obtained by the CONDON package must be truncated to the $|m_J\rangle$ components of the ground $J$ multiplet and then renormalized. Each eigenstate –either truncated and then renormalized or not- is written from left to right in the ordered basis set given by $|-J\rangle$ (leftmost coefficient),…, $|+J\rangle$ (rightmost coefficient), where $J$ is the ground electron spin quantum number.

The following row contains the initial population, at time $t=0$, of the lowest $2J+1$ eigenstates. These initial populations are accordingly ordered with the corresponding $2J+1$ eigenstates. Each initial population must be a number between 0.0 and 1.0. The summation of all initial populations must be equal to 1. In our case study, the initial population is all in the truncated eigenstate with $\langle J_z \rangle = -3.99$. For the remaining rows, see the section "Determination of relaxation pathways and identification of vibrations promoting relaxation".

rates.f:

The variables iorb, icpr, iram are switches to turn on or to turn off the calculation of the time evolution of magnetization by using the Orbach transition rates (output: rates.orbach.out), the calculation of the relaxation pathways and determination of the involved vibrations (output: rates.out), and the calculation of the time evolution of magnetization by using the second-order Raman transition rates (output: rates.raman.out), respectively. If one wants to set icpr = 1, then it is mandatory to set iorb = 1.

The variable temp is the working temperature (K), while sigma (cm$^{-1}$) is the Gaussian width employed in the transition rates (see section "Transition rates"). The number of time points to calculate the magnetization and to show both in rates.orbach.out and in rates.raman.out is npmag. The variables tim (s) and tfm (s) are the initial and final time points, respectively. Depending on which is the time scale where magnetization decays, one will have to change tim and tfm until detecting the magnetization decay with time. The number of vibrational modes is nmodos, dj is the numerical value of the ground electron spin quantum number $J$, and idtot is $2J+1$.

The following variables are thresholds which are necessary to take some important decisions. As it is known (see section "Resolution of the master equation"), one of the eigenvalues of the master matrix is strictly zero. Because of computational numerical noise and since computers work with finite precision arithmetic, the smallest eigenvalue is not zero (see description of the output rates.out below). The numerical value of this eigenvalue may be taken as a measure of the computational numerical noise. Since this smallest eigenvalue must be zero, we give the thresholds eot (s$^{-1}$) and ert (s$^{-1}$) a value a bit above the absolute value of this smallest eigenvalue. The threshold eot works when using the Orbach transition rates, while ert works in case of using the second-order Raman transition rates. One can first run a rates.f calculation, check the eigenvalues and then decide a value for both eot and ert. The threshold edt (cm$^{-1}$) is used to decide whether the energies of two given eigenstates are different. For the description of pot and mct see the section "Determination of relaxation pathways and identification of vibrations promoting relatxation".

rates.out: this output is broken down into three sections: "Orbach mechanism", "Orbach relaxation pathway", and "Raman mechanism". These sections will appear or not depending on which switches iorb, icpr, iram are turned on. At the beginning of each section, one can read the working temperature.

In the sections "Orbach mechanism" and "Raman mechanism" one finds first the eigenvalues of the master matrix. These eigenvalues must be always real and negative (but maybe the smallest one which, as said above, is technically zero). They are printed as complex numbers to check whether the imaginary part is zero. Then, the eigenvectors of the master matrix are printed row-wise and appear in the same order as that of the corresponding eigenvalues. Let us recall that the master matrix is built in the ordered basis set given by the lowest $2J+1$ eigenstates calculated at the relaxed geometry as provided in the input rates.inp (see section "Resolution of the master equation"). Thus, the components of each eigenvector follow the same order as that of these lowest $2J+1$ eigenstates in the rates.inp input. Then, one finds the coefficients $\{\lambda_k\}_{k=1}^{2J+1}$, which are the solution of a linear equation system built from the initial populations (see section "Resolution of the master equation"). Each one of these coefficients appear in Eq. S6 along with its corresponding eigenvalue $-1/t_k$. In the rates.out output, these coefficients are shown in the same order as that of the corresponding eigenvalues. Then, the lowest $2J+1$ energies calculated at the relaxed geometry (provided in the rates.inp input) are printed, along with the $J_z$ expectation values of the corresponding eigenstates provided also in the rates.inp input. To end up these two sections, one finds the master matrix.

In the section "Orbach relaxation pathway" one first finds the re-casted Orbach master matrix, which is subsequently normalized in terms of percentage (see section "Determination of relaxation pathways and identification of vibrations promoting relaxation"). To end up, it is shown those vibrational modes that contribute to the Orbach transition rates in the re-casted Orbach master matrix above a given threshold. This threshold is provided in the rates.inp input (see section "Determination of relaxation pathways and identification of vibrations promoting relaxation"). First, each pair i -> f identifies the relevant Orbach transition rate which is the component of the re-casted Orbach master matrix in the column "i" and in the row "f" (see section "Resolution of the master equation"). After i -> f, those modes whose percentage contribution (in parenthesis) to the corresponding Orbach transition rate is above the given threshold are displayed.

## Fortran codes to calculate CFPs thermal evolution

In the folder Home-Made-Codes one finds the codes cfptemp.f and cfppert.f. The input is frrmcfpd2.inp in both cases (see the section "Fortran code to evaluate magnetic relaxation dynamics").

cfptemp.f: this code calculates the CFPs $A_k^q \langle r^k \rangle (T)$ in cm$^{-1}$ at a given temperature $T$ (see the file CFPs-vs-Temp). To set the temperature, open the code and change the variable temp (K), where the number of vibrational modes is also found. The output is cfptemp.out. In this output, the CFPs at the relaxed geometry, at T = 0 K and at the given temperature are printed. The library lapack is not required. To compile, we use the compiler gfortran and the command line: gfortran –o aa cfptemp.f (aa is the executable name).

cfppert.f: this code calculates the parameters $\Delta \left( A_k^q \langle r^k \rangle \right)_j (T)$ in cm$^{-1}$ (see Eq. 1 in main text) for all vibrational modes $j$ at a given temperature $T$ (see the folder cfppert-temp). To set the temperature, open the code and change the variable temp (K), where the number of vibrational modes is also found. The output is cfppert.out. In this output, the parameters $\Delta \left( A_k^q \langle r^k \rangle \right)_j (T)$ are printed for each vibrational mode. These parameters are printed in two columns. In the left column and from top to bottom, they appear as: (2,0), (2,1), (2,2), (4,0), (4,1), (4,2), (4,3), (4,4), (6,0), (6,1), (6,2), (6,3), (6,4), (6,5), (6,6). In the right column and from top to bottom they appear as: nothing(always 0.00000000), (2,-1), (2,-2), nothing(always 0.00000000), (4,-1), (4,-2), (4,-3), (4,-4), nothing(always 0.00000000), (6,-1), (6,-2), (6,-3), (6,-4), (6,-5), (6,-6). In these parenthesis, the first number is the script k = 2, 4, 6, while the second number is the script q = -k,…, +k. Now, we use cfppert.out as an input for the code cfptomat.f (which can also be found in the folder Home-Made-Codes) in order to generate the perturbing Hamiltonians $\hat{H}_j$ that appear in the input matrix.inp (see section "Fortran code to evaluate magnetic relaxation dynamics"). Inside the code cfptomat.f one finds the following variables: inn, which selects the magnetic metal ion; dj, which is the metal ground electron spin quantum number $J$, idtot, which is $2J+1$ and has also to be changed when needed in the function oplm and in the subroutine operators; nmod, which is the number of vibrational modes. The library lapack is not required. To compile, we use the compiler gfortran and the command lines: gfortran –o aa cfppert.f and gfortran –o aa cfptomat.f (aa is the executable name).

## Description of the REC (Radial Effective Charge) model, and generation of an initial guess for the effective charges and the effective radial distances

The REC model is an electrostatic *semi-empirical* model commonly used in molecular magnetism, which provides an estimation of the crystal field parameters (CFPs) and allows rationalizing the magnetic properties of a given *f*-block single-ion magnetic coordination compound.(ref. 18 in main text) From the calculated CFPs, the model gives the ground-*J* multiplet energy levels and their corresponding wave-functions as a linear combination of the different $m_J$ = -J,…, +J microstates. For that, the (crystallographic) atomic coordinates of the first coordination sphere are required as an input. The REC model is implemented in the portable *fortran* SIMPRE computational package.([4] and ref. 11 in main text) This code parameterizes the electric field effect produced by the coordinating ligands by using the Crystal Field Hamiltonian in Eq. S8, expressed in terms of the Stevens Equivalent Operators (SEOs)[5]:

$$\hat{H}(J) = \sum_{k=2,4,6} \sum_{q=-k}^{+k} B_k^q \hat{O}_k^q = \sum_{k=2,4,6} \sum_{q=-k}^{k} (1-\sigma_k) A_k^q \langle r^k \rangle \eta_k \hat{O}_k^q \quad \text{Eq. S8}$$

In Eq. S8, *k* is the order (also called rank or degree) and *q* is the range, which varies between *k* and –*k*, of the SEOs $\hat{O}_k^q$. These are defined by Ryabov in terms of the angular momentum operators $J_\pm$ and $J_z$,[6] where the components $\hat{O}_k^q(c)$ and $\hat{O}_k^q(s)$ correspond to the SEOs with q ≥ 0 and q < 0 respectively.[7] Note that all the Stevens CFPs $A_k^q \langle r^k \rangle$ are real, whereas the matrix elements of $\hat{O}_k^q$ (q < O) are imaginary. $\eta_k$ are the $\alpha$, $\beta$ and $\gamma$ Stevens coefficients for *k* = 2, 4, 6, respectively, which are tabulated and depend on the number of *f* electrons.[8] $\sigma_k$ are the Sternheimer shielding parameters of the 4*f* electronic shell,[7] and <$r^k$> are the expectation values of the radius k[th] power.[9]

In SIMPRE, the $A_k^q$ parameters are determined by the following relations (second relation when q>0, third relation when q<0):

$$A_k^0 = \frac{4\pi}{2k+1} \sum_{i=1}^{N} \frac{Z_i e^2}{R_i^{k+1}} Z_{k0}(\theta_i,\varphi_i) p_{kq}$$

$$A_k^q = \frac{4\pi}{2k+1} \sum_{i=1}^{N} \frac{Z_i e^2}{R_i^{k+1}} Z_{kq}^c(\theta_i,\varphi_i) p_{kq}$$

$$A_k^q = \frac{4\pi}{2k+1} \sum_{i=1}^{N} \frac{Z_i e^2}{R_i^{k+1}} Z_{k|q|}^s(\theta_i,\varphi_i) p_{k|q|}$$

Eq. S9

In Eq. S9, $R_i$, $\theta_i$ and $\varphi_i$ are the effective polar coordinates of the point charges, and $Z_i$ is the magnitude of the effective point charge, associated to the *i*-th donor atom with the lanthanide ion at the coordinate origin, *N* is the number of ligands; *e* is the electron charge, $p_{kq}$ are the prefactors of the spherical harmonics and $Z_{kq}$ are the tesseral harmonics expressed in terms of the polar coordinates for the i-th donor atom.

In the REC model, any given ligand is modeled as an effective point charge placed between the lanthanide ion and the ligand coordinating atom at a distance $R_{eff}$ from the lanthanide ion, which is smaller than the real lanthanide-ion-donor-atom distance ($r_i$).(ref. 18 in main text) To account for the effect of covalency, a radial displacement vector ($D_r$) is defined, in which the polar coordinate *r* of each coordinating atom is collectively varied as $R_{eff} = r_i - D_r$, and at the same time the charge magnitude ($Z_i$) is scanned in order to achieve a minimum deviation between the calculated and the experimental target property P (e.g. the ground-*J* multiplet energy levels), whereas $\theta_i$ and $\varphi_i$ remain constant (see Fig. S1).

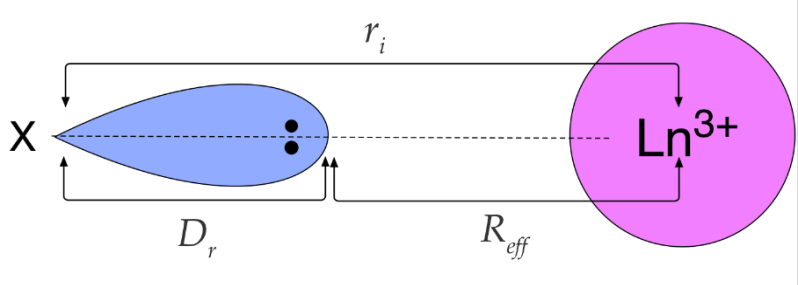

Figure S1: Lone electronic pair of a donor atom X oriented towards the nucleus of a trivalent lanthanide ion. The effective charge is located between the lanthanide ion and X at $R_{eff} = r_i - D_r$.

As a starting point for the fitting, we can estimate the effective distances of the coordinating atoms by using the following *semi-empirical* approximation for $D_r$:(ref. 19 in main text)

$$D_r \approx \left(\frac{N_L}{V_M}\right) \frac{1}{E_M(E_L - E_M)}$$ Eq. S10

In Eq. S10, $N_L$ is the coordination number of the lanthanide ion, $V_M$ is the valence of the lanthanide ion, and $E_M$ and $E_L$ are the Pauling electronegativity of the lanthanide ion and the donor atom, respectively. The effective charge $Z_i$ is estimated by using Eq. S11:

$$f_{X,CN} = D_r Z_i$$ Eq. S11

In Eq. S11, *f* is a factor that depends on the coordination number (CN) and on the coordinating atom (X).(ref. 19 in main text) The use of Eq. S11 is limited, since its use for a given system requires to know *f* in advance and, for that, one first has to know the REC parameters of a large enough set of coordination compounds with different lanthanide ion but with the same or similar ligands with the same coordination number.

In order to obtain the REC parameters ($D_r$ and $Z_i$) of the target compound, we need to vary both of them until a satisfactory fitting of a given property P –either experimental (e.g. spectroscopic energy levels, spectroscopically-determined CFPs, or magnetic properties) or calculated (e.g. via *ab initio* calculations)– is achieved. In the case of lanthanide single-ion coordination compounds, a fit of the ground-*J* multiplet energy levels will always be the desired option. Unfortunately, we cannot extrapolate this procedure to actinides, where the effects of excited multiplets are more important and thus are not negligible at all. In that case, in SIMPRE, one can fit the CASSCF or spectroscopically-determined CFPs using the REC model, and then obtain the energy level scheme by using the full Hamiltonian in the CONDON package.(ref. 20 in main text) The so-called *full model* therein considers inter-electronic repulsion, spin-orbit coupling, the ligand field potential, and, of course, both ground and excited J multiplets.[10] In the fitting procedures, we define the relative error *E* as:

$$E = \frac{1}{n} \sum_{i=1}^{n} \frac{\left[ P_{fit,i} - P_{ref,i} \right]^2}{\left[ P_{ref,i} \right]^2} \quad \text{Eq. S12}$$

In Eq. S12, $P_{ref}$ is the relevant property to fit and $P_{fit}$ is the best fit to $P_{ref}$, and *n* is the number of points used in the fiting.

## Projection on [U(Cp$^{ttt}$)$_2$]$^+$ of the CASSCF energies evaluated at the [Dy(Cp$^{ttt}$)$_2$]$^+$ experimental geometry

To determine the REC parameters that describe the crystal field produced by the two coordinating Cp$^{ttt}$ = {C$_5$H$_2$$^t$Bu$_3$-1,2,4} rings in [U(Cp$^{ttt}$)$_2$]$^+$, our starting point will be the CASSCF energy levels determined by Goodwin *et al.* at the experimental geometry of [Dy(Cp$^{ttt}$)$_2$]$^+$.(ref. 3 in main text) The experimental crystallographic coordinates of the first coordination sphere of [Dy(Cp$^{ttt}$)$_2$]$^+$ are used as an input in SIMPRE, and the two REC parameters are varied to fit the calculated CASSCF energy levels at the experimental geometry of [Dy(Cp$^{ttt}$)$_2$]$^+$. By using the REC model, the best fit, with an error of *E* = 0.03 % (Eq. S12), results in $D_r$ = 1.313 Å and $Z_i$ = 0.068. The calculated ground *J* multiplet energy levels ($E_{fit}$) by SIMPRE with these REC parameters are compared with the CASSCF ones ($E_{ref}$) in Table S2.

Table S2: Ground-*J* multiplet Kramers doublets determined by CASSCF ($E_{ref}$) and by the REC model ($E_{fit}$) for [Dy(Cp$^{ttt}$)$_2$]$^+$. $\Delta E = |E_{ref} - E_{fit}|$. Relative errors are < 2.6 %.

| $E_{ref}$–CASSCF (cm$^{-1}$) | $E_{fit}$–REC (cm$^{-1}$) | $\Delta E$ (cm$^{-1}$) |
|---|---|---|
| 0 | 0 | – |
| 488.6 | 480.7 | 7.9 |
| 771.0 | 775.6 | 4.6 |
| 956.5 | 980.9 | 24.4 |
| 1122.2 | 1145.4 | 23.2 |
| 1277.5 | 1280.5 | 3.0 |
| 1399.3 | 1365.4 | 33.9 |
| 1476.1 | 1465.4 | 10.7 |

Subsequently, we apply these calculated REC parameters to the DFT-relaxed coordinates of the coordinating atoms in [U(Cp$^{ttt}$)$_2$]$^+$. This target compound has identical ligands as [Dy(Cp$^{ttt}$)$_2$]$^+$ and only the metal ion is different. This allows us to transfer the REC parameters from [Dy(Cp$^{ttt}$)$_2$]$^+$ to [U(Cp$^{ttt}$)$_2$]$^+$ as demonstrated in several works.([11,12] and refs. 17, 19 in main text) The input coordinates of the relaxed positions of the coordinating atoms in [U(Cp$^{ttt}$)$_2$]$^+$ (*simpre.dat* file) and the calculated CFPs (*simpre.out*) are reported in Table S3 and Table S4, respectively. This procedure is systematically repeated for each distorted geometry along each vibrational mode, and the corresponding set of CFPs is obtained by performing a millisecond calculation in SIMPRE.

Table S3: Relaxed input coordinates of the coordinating atoms in [U(Cp$^{ttt}$)$_2$]$^+$ after applying $D_r$ = 1.313 Å to the radial coordinate and using a magnitude charge of $Z_i$ = 0.068.

| Label | $R_{eff}$ (Å) | $\theta$ (°) | $\phi$ (°) | $Z_i$ |
|---|---|---|---|---|
| C1 | 1.3017302 | 16.9734964 | 170.7902831 | 0.06806 |
| C2 | 1.3061774 | 15.4279100 | 349.0185755 | 0.06806 |
| C3 | 1.3907467 | 39.0491022 | 39.7965899 | 0.06806 |
| C4 | 1.4903647 | 49.2710181 | 80.8124802 | 0.06806 |
| C5 | 1.3744865 | 40.4952024 | 122.0849379 | 0.06806 |
| C6 | 1.3016208 | 163.0437971 | 63.0361018 | 0.06806 |
| C7 | 1.3061446 | 164.5540736 | 244.7622728 | 0.06806 |
| C8 | 1.3908882 | 140.9353178 | 194.0198766 | 0.06806 |
| C9 | 1.4905911 | 130.7250294 | 153.0093834 | 0.06806 |
| C10 | 1.3745142 | 139.5118956 | 111.7435627 | 0.06806 |

Table S4: Calculated CFPs for the [U(Cp$^{ttt}$)$_2$]$^+$ DFT-relaxed geometry in Stevens ($A_k^q \langle r^k \rangle$ and $B_k^q$) and Wybourne ($B_{kq}$) notation.

| k | q | $A_k^q \langle r^k \rangle$ (cm$^{-1}$) | $B_k^q$ (cm$^{-1}$) | $B_{kq}$ (cm$^{-1}$) |
|---|---|---|---|---|
| 2 | 0 | 997.8 | -6.414 | 1995.6 |
| 2 | 1 | -1118.3 | 7.188 | -456.5 |
| 2 | -1 | -568.0 | 3.651 | -231.9 |
| 2 | 2 | -47.1 | 0.303 | -38.5 |
| 2 | -2 | -64.4 | 0.414 | -52.6 |
| 4 | 0 | 262.0 | -0.076 | 2095.8 |
| 4 | 1 | -623.1 | 0.181 | -557.3 |
| 4 | -1 | -316.2 | 0.092 | -282.8 |
| 4 | 2 | -52.3 | 0.015 | -66.1 |
| 4 | -2 | -71.1 | 0.021 | -89.9 |
| 4 | 3 | -36.9 | 0.011 | -12.5 |
| 4 | -3 | -228.9 | 0.067 | -77.4 |
| 4 | 4 | -30.2 | 0.009 | -28.9 |
| 4 | -4 | 95.2 | -0.028 | 91.0 |
| 6 | 0 | 67.4 | -0.003 | 1079.1 |
| 6 | 1 | 823.0 | -0.031 | 1016.0 |
| 6 | -1 | 419.4 | -0.016 | 517.7 |
| 6 | 2 | 374.3 | -0.014 | 584.5 |
| 6 | -2 | 512.9 | -0.019 | 800.9 |
| 6 | 3 | -202.4 | 0.008 | -158.0 |
| 6 | -3 | -1254.1 | 0.048 | -979.1 |
| 6 | 4 | -184.9 | 0.007 | -117.9 |
| 6 | -4 | 581.4 | -0.022 | 370.6 |
| 6 | 5 | -98.3 | 0.004 | -29.9 |
| 6 | -5 | 98.6 | -0.004 | 30.0 |
| 6 | 6 | -50.6 | 0.002 | -53.3 |
| 6 | -6 | 17.1 | -0.001 | 18.0 |

Finally, the CFPs in Wybourne notation of the [U(Cp$^{ttt}$)$_2$]$^+$ DFT-relaxed geometry are used as an input in the CONDON package to determine the equilibrium electronic structure of [U(Cp$^{ttt}$)$_2$]$^+$, reported in Table S5. The wave-functions as determined by CONDON are expressed as a linear combination of the several J multiplets (both ground and excited). We truncate them to the ground J multiplet and then renormalize the resulting expression.

Table S5: Ground-*J* multiplet energy level scheme (Kramers doublets in cm$^{-1}$) and |m$_J$> contributions (of the same ground-*J* multiplet) to the wave-functions calculated for the relaxed geometry of [U(Cp$^{ttt}$)$_2$]$^+$.

| *E* (cm$^{-1}$) | Wave-function |
|---|---|
| 0 | 81.5% |+9/2> + 12.2% |+3/2> + 5.4% |+5/2> |
| 0 | 81.5% |-9/2> + 12.2% |-3/2> + 5.4% |-5/2> |
| 159.3 | 40.7% |-5/2> + 17.7% |-3/2> + 17.6% |+1/2> + 13.5% |-7/2> + 5.6% |+3/2> + 3.4% |+9/2> |
| 159.3 | 40.7% |+5/2> + 17.7% |+3/2> + 17.6% |-1/2> + 13.5% |+7/2> + 5.6% |-3/2> + 3.4% |-9/2> |
| 331.0 | 30.2% |+3/2> + 18.5% |+1/2> + 18.2% |-5/2> + 12.0% |-7/2> + 7.9% |-1/2> + 5.0% |+9/2> + 3.1% |+7/2> + 2.2% |-3/2> + 1.9% |-9/2> |
| 331.0 | 30.2% |-3/2> + 18.5% |-1/2> + 18.2% |+5/2> + 12.0% |+7/2> + 7.9% |+1/2> + 5.0% |-9/2> + 3.1% |-7/2> + 2.2% |+3/2> + 1.9% |+9/2> |
| 499.1 | 26.9% |-1/2> + 25.6% |-5/2> + 15.7% |+7/2> + 12.4% |-7/2> + 10.3% |-3/2> + 5.8% |-9/2> + 1.8% |+5/2> + 1.1% |+1/2> |
| 499.1 | 26.9% |+1/2> + 25.6% |+5/2> + 15.7% |-7/2> + 12.4% |+7/2> + 10.3% |+3/2> + 5.8% |+9/2> + 1.8% |-5/2> + 1.1% |-1/2> |
| 638.3 | 42.6% |+7/2> + 21.5% |+3/2> + 18.0% |+1/2> + 9.5% |-1/2> + 4.2% |+5/2> + 2.1% |-5/2> + 1.6% |+9/2> |
| 638.3 | 42.6% |-7/2> + 21.5% |-3/2> + 18.0% |-1/2> + 9.5% |+1/2> + 4.2% |-5/2> + 2.1% |+5/2> + 1.6% |-9/2> |